\documentclass[12pt]{article}

\usepackage{amsmath,amsfonts}
\usepackage{latexsym}
\usepackage{amssymb}
\usepackage{epsfig,psfrag,cite}
\usepackage{stmaryrd}

\makeatletter   
\@addtoreset{equation}{section}   
\makeatother

\newcommand{\be}{\begin{equation}}  
\newcommand{\ee}{\end{equation}}  
\newcommand{\bea}{\begin{eqnarray}}  
\newcommand{\eea}{\end{eqnarray}}  
\newcommand{\Tr}{\operatorname{Tr}}

\newcommand{\g}{\mathfrak g}
\newcommand{\h}{\mathfrak h}
\newcommand{\Li}{\operatorname{Li}}
\newcommand{\floor}[1]{\left\lfloor #1\right\rfloor}
\newcommand{\vev}[1]{\left\langle #1\right\rangle}

\newcommand{\rep}[1]{{\bf #1}}
\newcommand{\crep}[1]{{\bf \overline{#1}}}
\newcommand{\nn}{\nonumber}
\newcommand\T{\rule{0pt}{2.6ex}}
\newcommand\B{\rule[-1.2ex]{0pt}{0pt}}

\begin{document}

\thispagestyle{empty}
\vspace*{.5cm}
\noindent

\vspace*{1.9cm}

\begin{center}
{\Large\bf Anomalies on Six Dimensional Orbifolds}
\\[2.3cm]
{\large G.~von Gersdorff}\\[.5cm]
{\it Department of Physics and Astronomy, Johns Hopkins University, 
3400 N Charles Street, Baltimore, MD 21218}
\\[.4cm]
{\small\tt (\,gero@pha.jhu.edu\,)}
\\[1.3cm]

{\bf Abstract}\end{center} We consider localized anomalies in six
dimensional $Z_n$ orbifolds.  We give a very simple expression for the
contribution of a bulk fermion to the fixed point gauge anomaly that
is independent of the order $n$ of the orbifold twist. We show it can
be split into three terms, two of which are canceled by bulk fermions
and Green-Schwarz four forms respectively. The remaining anomaly
carries an integer coefficient and hence can be canceled by localized
fermions in suitable representations. We present various examples in
the context of supersymmetric theories. Also we point out that the
six-dimensional gravitational symmetries generally have localized
anomalies that require localized four-dimensional fields to transform
nontrivially under them.

\newpage

\section{Introduction}

The fermionic quantum numbers of the Standard Model (SM) clearly hint
towards an embedding of the latter into a gauge theory of a simple
group such as $SU(5)$ or $SO(10)$, dubbed Grand Unified Theory
(GUT). Moreover supersymmetry can both control the smallness of the
electroweak scale as well as unify the three gauge couplings of the SM
with astonishing accuracy at a scale not far below the Planck
scale. String and $M$-theory typically predict gauge groups that
contain these groups as subgroups and also possess a large number of
supersymmetric vacua, some of which might yield the minimal
supersymmetric SM (MSSM) at low energies. However, they require the
presence of extra dimensions which have to be compactified in order to
make them invisible in our low energy world. A popular choice of
geometry is an orbifold \cite{Dixon:1985jw} as it also provides a
mechanism to break the extended supersymmetry, produce chirality and
reduce the larger gauge symmetry to the SM or its simplest GUT
extensions. From a field theoretic point of view, these are effective
supergravity theories with all but four of its 10 or 11 dimensions
compactified on flat manifolds with isolated singularities.  Although
there are only few supergravity limits of string theory, the number of
possible orbifolds is huge, and looking for the MSSM in this vast
``landscapes'' of vacua might be a hard task\footnote{See
Refs.~\cite{Ibanez:1987dw} for reviews and
Refs.~\cite{Buchmuller:2005jr} for recent efforts in this direction.}.
It is however quite conceivable that some of the extra dimensions are
larger than others and intermediate models with effectively fewer
extra dimensions could be realized in nature. In view of this, a lot
of effort has been made to construct five and six dimensional models
and break the GUT group by orbifolding down to the SM
\cite{Kawamura:2000ev,
Altarelli:2001qj,
Hebecker:2001jb,
Asaka:2001eh,
Hall:2001xr,
Hall:2001zb,
Watari:2002tf,
Hebecker:2003jt,
Hebecker:2003we}.

An important consistency requirement of any gauge theory is that the
underlying local symmetries are free of anomalies, i.e.~they should
not be spoiled by quantum effects. In this paper, we will only be
concerned with perturbative anomalies, related to gauge
transformations that can be continuously connected to the identity.
Anomalies generally signal that the fermionic measure of the path
integral is non invariant (although the action is) and are a one loop
effect.  While string theory has no anomalies by construction, in
field theory this is an additional independent restriction for model
building. In smooth compactifications, the anomaly freedom of the
higher dimensional bulk theory guarantees that the low energy theory
is anomaly free as well. Orbifolds, on the other hand, involve
singularities such as boundaries and conical defects, and even though
the bulk theory is anomaly free, anomalies localized on these lower
dimensional ``branes''\footnote{Throughout this paper we will use the
terms brane, (orbifold) singularity and fixed point interchangeably.}
can occur~\cite{ Arkani-Hamed:2001is,
GrootNibbelink:2002qp,
Gmeiner:2002es,
Asaka:2002my,
Scrucca:2002is,
vonGersdorff:2003dt,
Scrucca:2004jn}.
In the limit where the volume of the orbifold is
taken to zero, the sum of the localized anomalies should match the
anomaly as calculated from the zero modes of the theory. Although the
absence of any localized anomaly in particular implies a consistent
zero mode spectrum, the reverse statement is generally not true.  It
is important to notice that the occurrence of the localized anomaly is
an entirely local phenomenon, i.e.~it only depends on the physics in
the immediate neighborhood of the brane. Consequently it should be
canceled locally, i.e.~by the introduction of matter localized at
the fixed point. There is however one important exception to this
rule. As it is well known from nonsingular spaces, one can sometimes
cancel the fermion anomaly by so-called Green-Schwarz $p$-form fields
that have anomalous couplings to the gauge fields at tree level. An
extension of this mechanism to orbifolds has been proposed in
Refs.~\cite{Horava:1995qa,
Scrucca:2002is,Gmeiner:2002es,vonGersdorff:2003dt,Scrucca:2004jn}
which due to its close relation with \cite{Callan:1984sa} is sometimes
referred to as the inflow mechanism. This name stems from the fact
that the Green-Schwarz fields are bulk objects, whose anomaly can
produce inflow from the bulk to the brane that can cancel the
localized anomaly.  However, this inflow takes on a particular form
that in general does not match the localized anomaly from the
fermions. Moreover, even if cancellation is achieved at one fixed point,
generically other inflow at different fixed points is created. A
necessary though not sufficient condition for this mechanism to cancel
all localized anomalies is that the sum of all fixed point
anomalies (or equivalently the anomaly of fermion zero modes)
vanishes.

Localized anomalies on orbifolds have been examined in
Refs.~\cite{
Arkani-Hamed:2001is,
GrootNibbelink:2002qp,
Gmeiner:2002es,
Asaka:2002my,
Scrucca:2002is,
vonGersdorff:2003dt,
Scrucca:2004jn}.
The purpose of this paper is to gain more insight in the structure of
these anomalies in the six-dimensional case. In six dimensions (or
generically in two extra dimensions) the local geometry is that of a
conical singularity, and the gauge breaking at a particular fixed
point is entirely encoded in a gauge Wilson line $P_{\mathcal G}$
around the apex of the cone. The background gauge potential generating
this Wilson line is expected to be pure gauge in the bulk, and hence its
fields strength should be a distribution supported at the fixed point:
\be
\langle F\rangle=-F_0\, \delta(y)\,\qquad P_{\mathcal G}=\exp(i F_0)\,.
\label{background}
\ee
As such, one would naively expect that if one plugs the nontrivial
background gauge potential generating this Wilson line into the
formula for the 6d bulk anomaly of a 6d chiral fermion, one should be
able to extract the anomaly at this conical singularity.  For instance,
the (irreducible) 6d gauge anomaly becomes a 4d brane
anomaly\footnote{See Sec.~\ref{anomalies} for a short introduction to
the concept of anomaly polynomials.}
\be
\Tr F^4\to -\Tr  F_0 F^3 \delta(y)\,.
\label{eq1}
\ee
If this naive expectation were true, the cancellation of the bulk
anomaly would immediately lead to the cancellation of the localized
anomaly. We will show that this simple assumption has to be modified
in two important ways. Firstly, rather than $F_0$ it is $(\log
P_{\mathcal G})/i$ that has to be inserted in the six dimensional
anomaly. While this seems to be equivalent by means of
Eq.~(\ref{background}) it is so only integers due to the
multivaluedness of the logarithm.  As a consequence, 6d bulk anomaly
cancellation no longer implies 4d brane anomaly cancellation, but
merely states that the localized anomaly carries an integer
coefficient.\footnote{In constructions with a resolved orbifold
singularity\cite{Serone:2004yn}, Eq.~(\ref{eq1}) is correct as it
stands. The singularity is obtained as a limit of a smooth background
field configuration, which also creates the brane localized zero
modes.  The difference of $F_0$ and $(\log P_\mathcal G)/i$ 
takes care of the anomaly of the latter.}  This is very satisfactory,
as it allows one to cancel the localized anomaly with localized
fermions. The contribution of single bulk fermions to the localized
anomaly usually has a fractional coefficient which only converts into
an integer once all the anomaly free bulk fermion spectrum is summed
over.  The other modification we will find is that in general further
terms, independent of the Wilson line, may appear. We will show that
these always take the form that can be canceled by the orbifold
version of the Green-Schwarz mechanism. In summary, the brane anomaly
can be split into three parts, which are canceled by bulk fermions,
brane fermions and Green-Schwarz forms respectively.

Another type of anomaly we will be concerned with involves the
higher dimensional gravitational symmetries. Anomalies of general
coordinate invariance and local Lorentz symmetry are intimately
related and can lead to non-conservation of energy-momentum and/or to
non-symmetric energy momentum tensors
\cite{Alvarez-Gaume:1983ig,Bardeen:1984pm,Chang:1984ib}. At orbifold
fixed points, some of the 6d gravitational symmetries are projected
out and others survive.  The latter fall into two classes from a 4d
point of view: They are either 4d gravitational symmetries or simple
gauge symmetries, albeit with a gauge boson that is a composite of
scalars (i.e.~the vielbein). For instance, the local Lorentz
transformations generated by $\Lambda_{AB}(x,y)$ become 4d local
Lorentz transformations $\Lambda_{\alpha\beta}(x,0)$ or local $SO(2)$
transformations $\Lambda_{ab}(x,0)$ at the fixed point, while the
mixed quantities $\Lambda_{\alpha a}(x,0)$ etc.~are forced to vanish
due to the orbifold boundary conditions. Although these gauge
symmetries are clearly not dynamical, they can develop anomalies at
the fixed point even if their corresponding bulk anomalies are
canceled. From a 6d point of view, the quantum effective action is
not invariant under local Lorentz symmetry but picks up source terms
supported at the singularities. We will compute these violations of 6d
local Lorentz symmetry and general coordinate invariance from fermions
propagating in the bulk. Quite interestingly, it turns out that in
general localized states transforming under this $SO(2)$ are needed to
render the theory consistent.  We point out that from a string
theory point of view localized states are indeed expected to
transform under these symmetries and hence contribute to (and possibly
cancel) these anomalies.

The paper is organized as follows: In Sec.~\ref{group} we review basic
facts about the theory of Lie algebras and their breaking by inner
automorphisms. Sec.~\ref{anomalies} contains the principal part of
this paper. We start by reviewing how localized anomalies arise in
orbifolds and then move on to analyze in depth the 6d geometry
$\mathbb R^4\times\mathbb R^2/Z_n$ for both gauge and gravitational
anomalies. Some of the technicalities are deferred to appendices
\ref{heatkernel} and \ref{evalsums}.  In Sec.~\ref{compact} we rewiew
the construction of compact torus orbifolds. Emphasis is placed on a
description in terms of Wilson lines around the singularities
(downstairs approach) which allows direct application of the results
of the previous section to the compact case.  We also point out some
peculiarities of compactifications related to the gravitational
anomalies. In Sec.~\ref{examples} we calculate gauge anomalies in
supersymmetric models and illustrate the formalism and the results of
Sec.~\ref{anomalies} with various examples.

\section{Symmetry Breaking}
\label{group}

In this section we would like to gather some known facts about Lie
algebras and symmetry breaking on orbifold fixed points.  For an
introduction to Lie algebras see e.g.~\cite{Slansky:1981yr} or the
textbooks in Ref.~\cite{books}.

\begin{figure}[tbh]
\psfrag{1}[b]{\footnotesize 1}
\psfrag{2}[b]{\footnotesize 2}
\psfrag{3}[b]{\footnotesize 3}
\psfrag{4}[b]{\footnotesize 4}
\psfrag{5}[b]{\footnotesize 5}
\psfrag{6}[b]{\footnotesize 6}
\psfrag{7}[b]{\footnotesize 7}
\psfrag{8}[b]{\footnotesize 8}
\psfrag{a}[b]{\footnotesize $n-2$}
\psfrag{b}[b]{\footnotesize $n-1$}
\psfrag{c}[b]{\footnotesize $n$}
\begin{center}
\includegraphics
[width=.8\linewidth]
{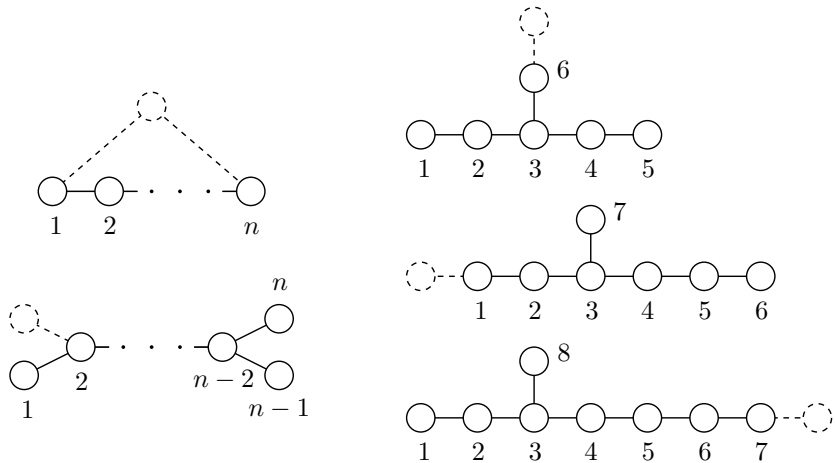}
\end{center}
\caption{\em Extended Dynkin diagrams of the simply laced Lie algebras. Left panel: $A_n=SU(n+1)$ (top) and $D_n=SO(2n)$ (bottom). Right panel (from top to bottom) $E_6$, $E_7$, $E_8$. The most negative roots $-\theta$ are displayed with dashed lines.}
\label{dynkin}
\end{figure}

The breaking of the gauge symmetry is best discussed in the
Cartan-Weyl basis of the corresponding Lie algebra $\g$.  For any $\g$,
one chooses a maximal subset of mutually commuting elements, the Cartan
subalgebra $\g_0$. The dimension of $\g_0$ is called the rank $r$ of
$\g$ and it is independent of the choice of $\g_0$. Each $H\in \g$ can
be simultaneously diagonalized in the adjoint representation,
\be
H_{adj} E_\alpha\equiv [H,E_\alpha]=\alpha(H) E_\alpha\,.
\label{cartan}
\ee
The quantities $\alpha$ are called roots. As is obvious from
Eq.~(\ref{cartan}), they are elements of the dual space $\g_0^*$ of
$\g_0$.  The set of all roots is called the root system; it contains
$\dim(\g)-r$ elements. As the nonzero eigenvalues are non degenerate, they are
in one-to-one correspondence with the generators $E_\alpha$.  The
elements of $\g_0$ together with the $E_\alpha$ span the whole algebra
$\g$.  We also define an inner product
\be
(H,H')=c\, \Tr_{adj} H H'\,,
\label{killing}
\ee
where $c$ is fixed in a moment. As Eq.~(\ref{killing}) is non
degenerate (in fact positive definite) it provides an isomorphism
between $\g_0$ and $\g_0^*$, and its inverse is a metric on root
space.  The root system of a general Lie algebra turns out to be
highly constrained and has lead to the famous complete classification
of all simple Lie algebras. Besides the four infinite series $A_r$,
$B_r$, $C_r$ and $D_r$ -- corresponding to $SU(r+1), SO(2r+1), SP(2r)$
and $SO(2r)$ respectively -- there are five exceptional algebras
$G_2$, $F_4$, $E_6$, $E_7$ and $E_8$.  One can choose a set of $r$
roots, $\{\alpha_{(i)}\}$, called the simple roots, with the remaining
roots being linear combinations of the simple roots
\be
\alpha = \pm \sum _i a^i \alpha_{(i)}\,,
\label{roots}
\ee
with the $a^i$ positive integers. Depending on the sign in
Eq.~(\ref{roots}) the roots are called positive or negative
respectively.  There is a unique positive root, the highest root
$\theta=c^i\alpha_{(i)}$ such that $c^i$ is greater than or equal to
the $a^i$ of all other positive roots.  The $c^i$ are called Coxeter
labels. The normalization of the Killing metric, i.e.~the constant $c$
in Eq.~(\ref{killing}), is now fixed by demanding that
\be
(\theta,\theta)=2\,.
\ee
We will from now on restrict ourselves to the $ADE$ algebras $A_r$,
$D_r$ and $E_r$, which have the property that all roots have length
squared two and the inner product of any two roots is either $0$ or
$\pm 1$ (for two simple roots only $0$ and $-1$ occur).  All
information of a Lie algebra is completely encoded in the simple roots
and can be neatly summarized in terms of Dynkin diagrams. Each node
stands for a simple root, and two nodes are joint by a single link if
their inner product equals $-1$. By adding to these diagrams the most
negative root $-\theta$ one obtains the so-called extended Dynkin
diagram which turns out to be rather useful in symmetry breaking.  The
extended Dynkin diagrams for the $ADE$ algebras are depicted in
Fig.~\ref{dynkin}.

The basis of simple roots viewed as a basis on $\g_0^*$ induces a
basis $\mu^{(j)}$ on $\g_0$ defined by
\be
\alpha_{(i)} \mu^{(j)}
={\delta_i}^j\,.
\label{fundw}
\ee
The $\mu^{(j)}$ are called fundamental weights.  
Owing to the
isomorphism induced by the metric, we will stop distinguishing between
$\g_0$ and its dual, writing any element in either basis
\be
\lambda = \lambda^i \alpha_{(i)}= \lambda_i \mu^{(i)}\,.
\ee
The coefficients $\lambda_i$ are called Dynkin labels. The Dynkin
labels of any root are also integers. Generalizing Eq.~(\ref{cartan})
we can define the roots for arbitrary representations, they are called
weights.  The importance of the basis Eq.~(\ref{fundw}) stems from the
fact that the Dynkin labels of all weights are integers (unlike the
roots, the weights expressed in the basis of simple roots are in
general not integers). 

We would like to break the gauge symmetry by a $Z_n$ inner
automorphism of $\g$, that is we would like to consider the map
$P_{\mathcal G}$ of the algebra onto itself:
\be
P_{\mathcal G}:T\rightarrow \exp\left( 2\pi i\, V_{adj} \right)T\,,
\qquad (P_{\mathcal G})^n=1\,,
\label{inner}
\ee
where $V$ is some element of $\g_0$ known as the shift vector. It is
 common to give $V$ in the basis Eq.~(\ref{fundw}), $V=V_i \mu^{(i)}$.
The generators which are left invariant by $P$ will form the algebra
$\h$ generating the unbroken gauge group $\mathcal H$. Clearly, each
element of $\g_0$ is of that type, so inner automorphisms do not break
the rank of the group. The generators $E_\alpha$ transform
non-trivially
\be
E_\alpha\to \exp(2\pi i\,V_i a^i) E_\alpha\,.
\ee
They will be projected out unless $V_i a^i$ is an integer.  In order
to get $(P_{\mathcal G})^n=1$ one demands
\be
n\, V^i a_i\in \mathbb Z\,,
\ee
which immediately leads one to 
\be
V_i=\frac{\nu_i}{n}\,,\qquad \nu_i\in \mathbb Z\,. 
\label{V}
\ee
Note that the $V^i$ are only defined modulo integers, so equivalent
shift vectors form a lattice.

We now would like to answer the two (complementary) questions: Which
subalgebras $\h\subset \g$ can be obtained in this way (and what is the
corresponding shift vector)? Given an arbitrary shift vector
satisfying Eq.~(\ref{V}), what is the resulting subalgebra $\h$?  To
answer the first question, it has been proposed \cite{Choi:2003pq} to
transform $V$ to a canonical form, by means of lattice transformations
and so-called Weyl reflections (the reflections of the root system on
hyperplanes perpendicular to a given root).  Using these symmetries
one casts $V$ into the form
\be
\nu_i c^i\leq n\,,\qquad(\nu_i\geq0)\,.
\label{canonical}
\ee
It is a simple exercise to specify all shift vectors satisfying
Eq.~(\ref{canonical}) once $\g$ and $n$ are specified.  Finding the
surviving subgroup defined by this canonical shift vector is now very
simple, first treat the case $\nu_i c^i< n$.  All simple roots
$\alpha_{(i)}$ with $\nu_i\neq0$ are projected out and the same is
true for any non-simple roots that contain one or more such simple
roots in its decomposition Eq.~(\ref{roots}).  The unbroken subgroup
is thus found by deleting the nodes of the Dynkin diagram with
$\nu_i\neq0$ and reading off the semisimple part. The remaining $U(1)$
generators are given by
\be
T^{(i)}=\mu^{(i)}\,.
\label{charges}
\ee
This choice gives integer charges to the $\mathcal H$-representations
occurring in the branching rule of the adjoint of $\mathcal G$.
Notice that these $U(1)$ generators are only orthogonal w.r.t.~the
Killing metric Eq.~(\ref{killing}) if they correspond to non-adjacent
nodes in the Dynkin diagram.  We will need the values for these $U(1)$
charges. They can in principle can be taken over from the literature
(see for instance \cite{Slansky:1981yr}) and renormalizing them such
that the smallest charge occurring in the branching rule of the
adjoint is $\pm 1$, which fixes the charge up to a sign.\footnote{An
explicit way to obtain the charge is as follows: the branching rules
for the irrep with highest weight $w$ contains the representation with
highest weight $w'$ which is simply $w$ with the $i^{\rm th}$ Dynkin
index removed.  In our normalization Eq.~(\ref{charges}) its charge is
$G_{ij}w^j$ ($G$ being the metric).  As an example, consider the
breaking of $SO(10)\to SU(5)\times U(1)$. This can be achieved by the
$Z_n$ shift vector $V=\mu_{(4)}/n=(00010)/n$. The representation with
highest weight $w=(00001)$ is the {\bf 16} which contains a $SU(5)$
representation with highest weight $w'=(0001)$ that is identified as
the $\crep 5$ of charge $G_{45}=3/4$. The charges of the remaining
representations in the branching differ from this by integer multiples
of $n$.}
%
%
The surviving subgroup in the case $\nu_i c^i=n$ is found by deleting
the corresponding nodes of the {\em extended} Dynkin diagram.  In particular, by deleting a single node, all maximal (regular) subalgebras can be obtained
\cite{Hebecker:2003jt}. 
%
%

Let us turn to the other question raised above, which subgroup is
obtained for an arbitrary $V$. One could find the corresponding Weyl
reflections and lattice shifts that transform $V$ into the canonical
form Eq.~(\ref{canonical}). However, this might be a very complicated
task. On the other hand, with the knowledge of the root system it is
quite easy to give a diagrammatic recipe that can be used.  The
complication for arbitrary shift vectors stems from the fact that
although it is obvious which simple roots are projected out, some
non-simple roots which are not just the linear combinations of the
surviving simple roots can be projected in.  One thus has to scan the
entire root system, find the surviving roots and reconstruct the
unbroken algebra. We would like to describe a simpe method that
simplifies this procedure.  Without loss of generality we assume the
shift vector to take the form $0\leq V_i<1$.  The set of simple roots
decomposes into
\be
\Sigma_{\g}=\Sigma_{\h_0}\cup \Sigma
\ee
where $\Sigma_{\h_0}$ contains the simple roots satisfying $V\cdot
\alpha_{(i)}=0$ and $\Sigma$ the simple roots that are projected out.
We would like to complete the set $\Sigma_{\h_0}$ to the set
$\Sigma_{\h}$, i.e.~we would like to find further surviving roots that
can serve as simple roots.  To this end one has to write down the root
system $\Gamma_\g$ in terms of the simple roots (the root systems are
well known and can be taken from the literature
\cite{Slansky:1981yr,books} or can be computed with the help of a
computer algebra system such as LiE \cite{lie}). Identify the set of
roots $\Gamma_\h$ that survive the projection and write any surviving
positive root $\beta\in\Gamma_\h$ as
\be
\beta=\gamma+\alpha\,,\qquad \alpha\in
\operatorname{span}(\Sigma_{\h_0})\,,
\quad
\gamma\in \operatorname{span}(\Sigma)
\ee
the set of surviving roots splits into equivalence classes labeled by
$\gamma$. In fact the roots in these equivalence classes are nothing
but the weights of the representations in the branching of the adjoint
in the breaking $\h\to\h_0$ 
There are thus unique heighest roots
\be
\beta_\gamma=\gamma+\alpha_\gamma
\ee
in each class. The new simple roots will be chosen from the set
$\{-\beta_\gamma\}$. In fact all one has to do is to consider the set
$\Gamma=\{\gamma\}$ and find a linear independent subset $\Gamma'$ such
that an arbitrary $\gamma$ is a linear combinations of the $\gamma'\in\Gamma'$
with positive integer coefficient.\footnote{The set $\Gamma'$
usually consists of those $\gamma'$ with small $V\cdot\gamma'$.} 
In the last step, draw the Dynkin diagram of the set
$\Sigma_\h=\Sigma_{\h_0}\cup \{-\beta_{\gamma'}\}$ and read off the
surviving algebra $\h$.

\begin{figure}[thb]
\begin{center}
\includegraphics[width=10cm]{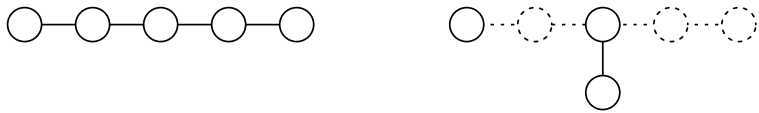}
\end{center}
\caption{\em Left: Dynkin diagram of $SU(6)$. Right: breaking pattern of the shift vector $V=(0,1/4,0,1/2,1/2)$ (see text).} 
\label{su6}
\end{figure}

Although this approach seems rather pedestrian, it turns out to be
quite efficient and is certainly simpler than identifying the Weyl
reflection that leads to the canonical form.  An example is in
order. Consider $A_5=SU(6)$ and the shift vector $V=(0,1/4,0,1/2,1/2)$.
We thus have 
\be
\Sigma_{\h_0}=\{\alpha_{(1)}\,,\ \alpha_{(3)}\}\,,\qquad
\Sigma=\{\alpha_{(2)}\,,\ \alpha_{(4)}\,,\ \alpha_{(5)}\}\,.
\ee
The complete root system of $A_r$ is given by
\be
\Gamma_{A_r}=\{\pm(\alpha_{(\ell)}+\alpha_{(\ell+1)}+\dots+\alpha_{(k)})\,,\ 0\leq\ell\leq k\leq r\}\,.
\ee
Besides the simple roots in $\Sigma_{\h_0}$, the only surviving
positive roots are $\beta_1=\alpha_{(4)}+\alpha_{(5)}$ and
$\beta_2=\alpha_{(3)}+\alpha_{(4)}+\alpha_{(5)}$. There is only one
equivalence class labeled by $\gamma=\alpha_{(4)}+\alpha_{(5
)}$ and the
highest root in this class is $\beta_2$. One thus finds the new set of
simple roots
\be
\Sigma_{\h}=\Sigma_{\h_0}\cup \{-\beta_2\}=
\{\alpha_{(1)}\,,\ \alpha_{(3)}\,,\ -\alpha_{(3)}-\alpha_{(4)}-\alpha_{(5)}\}\,.
\ee
From the original Dynkin diagram one reads off
\be
- \beta_2\cdot\alpha_{(1)}=0\,,\qquad - \beta_2\cdot\alpha_{(3)}=-1\,,
\ee
hence the new simple root is connected with a single line to
$\alpha_{(3)}$.  The final Dynkin diagram is that of $SU(3)\otimes
SU(2)$. Since this has rank three, there are two $U(1)$ factors.  The
example is displayed in Fig.~\ref{su6}.

\section{Fixed Point Anomalies}
\label{anomalies}

Orbifolds are obtained from smooth manifolds by modding out a discrete
(in general finite) symmetry group $\mathbb G$.  The points that are
left invariant by some $P\in \mathbb G$ are called the fixed points of
$P$, they form singular subspaces of lower dimension such as
boundaries and conical defects.  An anomaly-free theory in $d$
dimensions does not guarantee that an orbifolded version of that
theory is non-anomalous. In particular, higher dimensional fermions
generically contribute an anomaly supported at the fixed
points~\footnote{The factor of $i$ arises because we work in Euclidean
space. Eq.~(\ref{divergence}) is normalized such that $\mathcal A$ is real and
the Jacobian of the path integral equals $\exp[i\int A(\Lambda)]$.}
\be
\Lambda^A D_M j_A^M = -i\mathcal A(\Lambda) \delta(y,y_f)\,,
\label{divergence}
\ee
where the form of the localized anomaly $\mathcal A(\Lambda)$
corresponds to the dimension of the fixed point. To cancel the
anomaly, localized fermions have to be introduced.  Then contribution
to the anomaly from bulk fermions has been studied by many authors in
a variety of different models~\cite{ Arkani-Hamed:2001is,
GrootNibbelink:2002qp,
Asaka:2002my,
Scrucca:2002is,
vonGersdorff:2003dt,
Scrucca:2004jn}.
Using Fujikawas path integral method \cite{Fujikawa:1979ay}, in
Ref.~\cite{vonGersdorff:2003dt} the following compact expression for
the anomaly was derived:\footnote{To fully connect with
Ref.~\cite{vonGersdorff:2003dt} we note that we will focus on even $d$
(in particular $d$=6) in this paper. Furthermore we will choose our
orbifold twists to satisfy $Q_\psi=Q_{\bar\psi}$ in the language of
Ref.~\cite{vonGersdorff:2003dt}.}
\be
\mathcal A = -\lim_{s\to 0} \Tr \Lambda\, \Gamma Q_\psi 
\exp(s D^2\!\!\!\!\!\!/\ )\,.
\label{master}
\ee
Eq.~(\ref{master}) is nothing but the Jacobian of the gauge
transformation in the (Euclidean) path integral.  Here $\Lambda$ is
the generator of some (in general nonabelian) symmetry, $\Gamma$ is the $d$
dimensional chirality matrix (the analogue of $\gamma^5$ in four
dimensions) and $Q_\psi$ is the orbifold projector defined as
follows. Consider a chiral fermion in the $d$ dimensional theory that
satisfies the orbifold boundary condition
\be
\psi(x)= P_{\mathcal G}\otimes  P_{L}\, \psi(P^{-1} x)\,,
\label{constraint}
\ee
under the orbifold group element $P$. Here $P_{\mathcal G}$ and
$P_{L}$ define the action of the orbifold twist on the internal 
and Lorentz indices respectively.  $P_{\mathcal G}$ is the Wilson line
that -- if nontrivial -- will give rise to the orbifold breaking of
the gauge group.  By defining an analogous action $\hat P$ of the
orbifold group on function space via
\be
\langle x|\hat P |x' \rangle = \delta(x,P x')\,, 
\ee
we can define the action of $P$ on the field $\psi$ as $P_\psi=\hat
P\otimes P_{\mathcal G}\otimes P_{L}$. The orbifold conditions
$P_\psi\psi=\psi$ can now be solved explicitly in terms of the orbifold
projector $Q_\psi$
\be
\psi=Q_\psi \tilde \psi,\qquad Q_\psi\equiv \frac{1}{n}\sum_P P_\psi\,,
\label{proj}
\ee
where the sum goes over all $n$ orbifold group elements and $\tilde
\psi$ denotes an arbitrary unconstrained field.  The orbifold
projector acting on an $\tilde \psi$ yields a field $\psi$ satisfying
the constrained Eq.~(\ref{constraint}).  The expression
Eq.~(\ref{master}) is now a rather intuitive generalization of
Fujikawas result \cite{Fujikawa:1979ay}: Just insert the orbifold
projector in the trace to restrict it over fields that satisfy the
orbifold boundary condition.  One can show that $Q_\psi$ satisfies the
following properties:
\be
Q_\psi^\dagger=Q_\psi,\qquad Q_\psi^2=1,\qquad [Q_\psi,\Gamma]=0\,,
\ee
as well as
\be
[D\!\!\!\!/\,,Q_\psi]=0,\qquad [\Lambda,Q_\psi]=0\,.
\ee
Notice that the expression Eq.~(\ref{master}) reproduces the bulk
anomaly from the term in the sum in Eq.~(\ref{proj}) that corresponds
to $P=id$. Whenever one of the $P's$ possesses a fixed point the
corresponding term in $Q_\psi$ produces a fixed point anomaly.
Eq.~(\ref{master}) has been evaluated in the flat background in
\cite{vonGersdorff:2003dt}.  Details on the evaluation for the case of
curved backgrounds are provided in App.~\ref{heatkernel} (see also
Ref.~\cite{Scrucca:2004jn} for an alternative calculation using the
formalism of Ref.~\cite{Alvarez-Gaume:1983ig}).

Before we go on and evaluate the anomalies for the case of six
dimensions, we would like to comment on the form of the anomaly as
obtained from Eq.~(\ref{master}).  The anomaly calculated in this way
gives the so-called covariant anomaly.  It transforms covariantly
under gauge transformations but does not satisfy the Wess-Zumino
consistency condition
\be
\delta_{\Lambda_1}\mathcal A(\Lambda_2)-\delta_{\Lambda_2}\mathcal
A(\Lambda_1)=\mathcal A([\Lambda_1,\Lambda_2])\,,
\label{WZCC}
\ee
that follows by identifying the gauge variation of the effective
action $\Gamma[A_M]$ with the anomaly.  The fact that the covariant
anomaly fails to fulfill Eq.~(\ref{WZCC}) can be traced back to the
form of the regulator used in Eq.~(\ref{master}). In fact, by a
different choice of regulator \cite{Alvarez-Gaume:1985ex} one can
arrive at the so-called consistent form of the anomaly which does
satisfy Eq.~(\ref{WZCC}).  As a matter of fact, the covariant and
consistent currents whose divergences are proportional to the
corresponding anomaly are related by local counterterms (that is,
local functionals of the gauge and Lorentz connections). Their
transformation under the symmetry considered precisely gives
the difference between the covariant and consistent anomalies
\cite{Bardeen:1984pm}. As far as cancellation of anomalies is
concerned the two forms are equivalent.  For many purposes, including
Green-Schwarz anomaly cancellation, the consistent form of the anomaly
is more convenient. Explicit expressions for the consistent anomaly in
$d$ dimensions can be derived from the anomaly polynomial via the
Wess-Zumino descent equations. The anomaly polynomial $I$ is a formal
$d+2$ form which is a polynomial of order $d/2+1$ in the two-forms $F$
and $R$, the field strengths of the gauge and Lorentz connections. For
instance, in four and six dimensions, the anomaly polynomial for a
spin $1/2$ field reads:\footnote{All Products are to be read as wedge
products. The trace over the $R$'s is taken in the fundamental of
$SO(d)$ and all generators are hermitian.}
%
%
\be
I(R,F)= \frac{1}{3!(2 \pi)^2}\left(-
\frac{1}{8}\Tr F\Tr R^2+\Tr F^3
\right)\,,
\label{poly4}
\ee
\be
I(R,F)=\frac{1}{4!(2\pi)^3}\left(\frac{d_r}{240}\Tr R^4+
\frac{d_r}{192}(\Tr R^2)^2-\frac{1}{4}\Tr R^2\Tr F^2+\Tr F^4\right)\,,
\label{poly6}
\ee
where $d_r$ denotes the dimension of the representation labeled by $r$.  From
this, the anomaly is calculated by writing $I$ as the derivative of
the Chern-Simons form $\omega$, whose variation is the derivative of
the anomaly:
\be
I=d\omega\,\qquad \delta \omega=d\mathcal A\,.
\label{descent}
\ee
These relations are known as the Wess-Zumino descent equations.  The
condition that the total anomaly be zero is equivalent to the
vanishing of the total anomaly polynomial (that is the sum of the
contributions from different fields). Anomalies whose polynomial
factorizes as in the second and third term in Eq.~(\ref{poly6}) are
called reducible, otherwise they are called irreducible. Reducible
anomalies might be canceled by the Green-Schwarz mechanism
\cite{Green:1984sg} that involves $p$-form fields with anomalous tree
level couplings to the Chern-Simons forms.  As observed in
\cite{Horava:1995qa,
Scrucca:2002is,Gmeiner:2002es,vonGersdorff:2003dt,Scrucca:2004jn}, the
Green-Schwarz mechanism can also give contributions to the localized
anomaly, which is also known as the inflow mechanism. These inflow
contributions are always of the form
\be
I^{(p)}\wedge I'^{(q)}\wedge \delta^{(m)}\,,\qquad p+q+m=d+2\,,
\label{inflow}
\ee
where the $p$-form $I^{(p)}$ is as polynomial invariant under the full
$d$-dimensional symmetries, the $q$-form $I'^{(q)}$ is a polynomial
that is only invariant under the symmetries of the fixed point and
$\delta^{(m)}$ is the $\delta$ function peaked at the singularity
(here interpreted as an $m$-form). As we will see in the following,
part of the localized anomalies generated by bulk fermions are of that
type and hence can be canceled by an appropriate Green-Schwarz
contribution.

\subsection{Localized Gauge Anomalies in 6d Orbifolds}
\label{localized}

We will now consider the six dimensional orbifold $\mathbb R^4\times
\mathbb R^2/Z_n$ with arbitrary integer $n$. The $Z_n$ symmetry
consists of the rotation with angle $2\pi/n$ around the origin and the
geometry of the two extra dimensions is the (infinite) cone with
opening angle $2\pi/n$. Compactifications (i.e.~orbifolds of the type
$T^2/Z_n$ with possible Wilson lines) will be considered in
Sec.~\ref{compact}.  However, the local geometry of these spaces in
the vicinity of a fixed point is identical to the non-compact case and
the results of this section are directly applicable.  To define a
proper $Z_n$ action on the fermions, Eq.~(\ref{constraint}), we note
that $(P_L)^n=-1$ and hence we demand the same for the internal twist,
$(P_{\mathcal G})^n=-1$. Notice that with $\mathcal G$ we refer to the
entire internal symmetry group that besides the local symmetries can
contain global ones such as flavour or $R$-symmetries. Correspondingly
the twist $P_{\mathcal G}$ has to be viewed as a product of several
factors which, in particular, contains the gauge twist in the
appropriate representation.  The contribution to the gauge anomaly of
a 6d spin $1/2$ fermion with positive chirality\footnote{For negative
6d chirality an additional sign has to be included.} at a $Z_n$
orbifold fixed-point can be written as
\be
\mathcal A_{\mathcal G}=\frac{1}{(2 \pi)^3} 
\Tr (\Lambda F_0 A_2)\,\delta(y)\,,
\label{anom}
\ee
\be
A_2=\frac{1}{8}\epsilon^{\alpha\beta\gamma\delta}\left(\frac{1}{24}R_{\alpha\beta AB}{R_{\gamma\delta}}^{AB}-F_{\alpha\beta} F_ {\gamma\delta}
\right)\,,
\label{A2text}
\ee
\be
F_0=\frac{i\pi}{n} \sum_{k=1}^{n-1}\frac{1}{\sin(\pi k/n)} 
(P_{\mathcal G})^k\,.
\label{F0}
\ee
Some details of the derivation of Eq.~(\ref{result}) from
Eq.~(\ref{master}) are presented in App.~\ref{heatkernel}.  Here $F_0$
can be seen to originate from the projector $Q$ in Eq.~(\ref{master})
once the functional and $\Gamma$ traces are performed and the bulk
anomaly is separated. Notice that the $\delta$-function singles out
the 4d field strengths of $\mathcal H\subset \mathcal G$ and
$SO(4)\otimes SO(2)\subset SO(6)$ respectively.

It will be more convenient for our purposes to convert the anomaly
Eq.~(\ref{anom}) first into the consistent form\footnote{In fact, all
we have to do is to add a Bose symmetrization factor $1/3$ to the
linearized terms of the cubic gauge anomaly. The nonlinear terms are
then completely fixed by the form of the consistent anomaly.} and
subsequently deduce the anomaly polynomial from which the anomaly
descends via Eq.~(\ref{descent}). The result is
\be
-\frac{1}{3!(2\pi)^3}\left( \Tr F_0 F^3 - 
\frac{1}{8} \Tr R^2 \Tr F_0 F\right)\delta(y)\,.
\label{polyorb}
\ee
Let us stress here that Eq.~(\ref{polyorb}) is truly a four
dimensional anomaly since the matrix $F_0$ is constant on each
irreducible representation of $\mathcal H$.  However,
Eq.~(\ref{polyorb}) suggests that we interpret $F_0$ as some kind of
background field strength supported at the singularity\footnote{Here
we interpret the $\delta$ function as a two form.}
\be
\vev F=-F_0\,\delta(y)\,.
\label{vevF}
\ee
Indeed, by replacing $F\to \vev F+F$ in the 6d polynomial
Eq.~(\ref{poly6}) we precisely\footnote{Notice that there is a
combinatorial factor of 4 in the pure gauge anomaly and a
corresponding factor of 2 in the mixed one. Also note that $\vev
F\wedge \vev F=0$ as $\delta\wedge \delta=0$. }  arrive at
Eq.~(\ref{polyorb}).  In other words there should be some gauge
connection $\vev A$ that is pure gauge in the bulk (so that its fields
strength $\vev F$ is supported at the fixed point) and that generates
the Wilson line $P_{\mathcal G}$ encircling the singularity.
Superficially, the definition Eq.~(\ref{F0}) seems to have little to
do with such a background field. However, one can actually perform the
finite sum in Eq.~(\ref{F0}) analytically. We present the calculation
in App.~\ref{evalsums}, the result is
\be
F_0= -i \log(-P_{\mathcal G})\,.
\label{result}
\ee
This is quite remarkable as it implies that the Wilson line generated
by gauging away the background field  Eq.~(\ref{vevF}) is
\be
\exp (-i \oint \vev A) = \exp (-i\!\int\!\vev F)=-P_{\mathcal G}\,,
\ee
i.e.~ the orbifold twist around the singularity (up to a
sign). However there are a few points worth observing. Firstly, the
quantity $F_0$ is not really Lie algebra valued for general
$P_\mathcal G$. The reason is the multivaluedness of the logarithm in
Eq.~(\ref{result}). As we will see below, an important consequence of
this is that after imposing cancellation of the 6d irreducible gauge
anomaly, the remaining irreducible 4d gauge anomaly has an integer
coefficient.  This is essential if localized fermions are to cancel
this anomaly.  Secondly, $P_{\mathcal G}$ might also contain global
symmetries. They can formally be promoted to local symmetries with a
fixed background field configuration. The last point concerns the sign
inside the logarithm in Eq.~(\ref{result}). As we will also see below,
this generates a term in the anomaly that can be canceled by the
Green-Schwarz mechanism.
As a matter of fact, this additional sign is actually absent if one
considers an additional $Z_2$ twist in the Lorentz bundle. Notice that
for particles of integer spin an angle of rotation of $2\pi/n$ is
indistinguishable from $2\pi/n+2\pi$. For fermions however, these two
rotations differ by an overall sign and hence instead of $P_L$ one can
as well consider $\tilde P_L=-P_L$.  For odd $n$ one has $(\tilde
P_{L})^n=+1$ and implementing this twist in the orbifold symmetry then
implies $(P_{\mathcal G})^n=+1$.  It is easy to evaluate the
corresponding sum with this modified $P_L$. We again find
Eq.~(\ref{anom}) but now $F_0$ is given by
\be 
F_0= -i \log(+P_{\mathcal G})\,.
\label{resultodd}
\ee

Let us now focus on the pure gauge anomaly and specialize to the case
where the Wilson lines are given by inner automorphisms\footnote{For
outer automorphisms, restrictions on the matter representations apply
(see, e.g.~Ref.~\cite{Hebecker:2001jb}) which makes the general
treatment more involved.} as described in Sec.~\ref{group}.  Let us
first get rid of the sign inside the $\log$ in Eq.~(\ref{result}) by
writing
\be
\log(-P_{\mathcal G})=\log'(P_{\mathcal G})-i\pi\,,
\label{log}
\ee
where the branch cut of $\log'$ lies just below the positive real axis.
Applying Eq.~(\ref{log}) to Eq.~(\ref{result}), we see that we are
left with two contributions 
\be
-\frac{1}{2\pi}\Tr \bigl[F_0 F^3\bigr]=-\frac{1}{2\pi i}
\Tr\bigl[\log'(P_{\mathcal G}) F^3]
+\frac{1}{2}\Tr \bigl[F^3\bigr]\,.
\label{split}
\ee
The second term is independent of the Wilson line.  It is the product
of a $\mathcal G$ invariant polynomial times a $\delta$-function
two-form, i.e.~it takes the form of Eq.~(\ref{inflow}) with
$q=0$. Consequently, it can be canceled by anomaly inflow from the
bulk. 
The first term in Eq.~(\ref{split}) 
\be
\tilde I\equiv-\frac{1}{2\pi i}\Tr \bigl[ \log'(P_{\mathcal G})F^3\bigr]\,,
\label{nonuniversal}
\ee
defines a localized anomaly that cannot in general be canceled by the
inflow mechanism. The localized anomaly in the form
Eq.~(\ref{nonuniversal}) is the starting point for a detailed
evaluation. This is best described with concrete examples and we
postpone it to Sec.~\ref{examples}.  For the remainder of this section
we want to prove an important theorem.  Using the form of the inner
automorphism Eq.~(\ref{inner}) we rewrite Eq.~(\ref{nonuniversal}) as
\be
\tilde I=\Tr \bigl[ -VF^3+\floor{V}\!F^3\bigr]\,
\label{nonuniversal2}
\ee
Recall that the shift vector $V$ is an element of the Cartan
subalgebra; here it has to be interpreted in the representation of the
fermion.  In Eq.~(\ref{nonuniversal2}) we have defined the floor function
$\floor{x}$ as the closest integer $\leq x$. First consider the case
where the inequality in Eq.~(\ref{canonical}) holds, i.e.~the highest
root is projected out by the inner automorphism. In this case $V$
corresponds to a $U(1)$ factor of $\mathcal H$.  The charges $q_{\rep r}$ of the irreducible $\mathcal H$ representations labeled by
$\rep{r}$ are normalized as described in Sec.~\ref{group}.  Notice
that the first term in Eq.~(\ref{nonuniversal2}) actually takes on the
form of a 6d bulk anomaly. Its irreducible part has to be canceled
with other bulk fermions, while its reducible part
\be
\sim \Tr (V F) \Tr (F^2)=\sum_{\rep r} q_{\rep r}\, F_{U(1)} \Tr_{\rep r} F^2
\label{mixed}
\ee
turns into a 4d reducible anomaly (i.e.~a mixed $U(1)$ nonabelian one)
as the first trace projects onto the $U(1)$ generators. The remaining
irreducible 4d anomaly is thus
\be
\tilde I=
\sum_{\rep r}\floor{q_{\rep r}/n} \Tr_{\rep r} \bigl[ F^3\bigr]\,. 
\label{integer}
\ee
The sum goes over all irreducible $\mathcal H$ representations as
determined from the branchings of the corresponding irreps of
$\mathcal G$ in the bulk. We see that the coefficients of the
irreducible fixed point anomaly are integers. This is indeed very
satisfactory. It guarantees that we can cancel these irreducible
anomalies by introducing brane localized fermions in appropriate
representations. Had we been left with any non-integer coefficients,
there would be no way of obtaining a consistent theory at the fixed
point. With the information from Eq.~(\ref{integer}) it is now possible to
obtain predictions on the localized fermion content.  For the modified
Lorentz twist $\tilde P_L$ the following simple modifications apply:
In Eq.~(\ref{nonuniversal}) $\log'$ is replaced by $\log$ and the
arguments of the floor functions in Eqns.~(\ref{nonuniversal2}) and
(\ref{integer}) are shifted by 1/2.

The case in which the equality in Eq.~(\ref{canonical}) holds and one
encounters the enhanced symmetry is more involved.\footnote{In
particular, $V$ and $\floor V$ are no longer separately constant on
each irreducible representation of $\mathcal H$.} The unbroken group
$\mathcal H$ contains a simple factor, $\mathcal H_\theta$ having the
root $-\theta$ as one of its simple roots. The localized anomaly of
the other group factors can actually be calculated as above by
formally considering the branching of $\mathcal H_\theta\to\mathcal
H'_\theta\otimes U(1)$. It remains to calculate the anomaly of
$\mathcal H_\theta$. Note that in 4d only the $SU(N)$ groups have
anomalies.  If $N>3$ one can calculate the anomaly of $\mathcal
H_\theta'=SU(N-1)$ as above which is an integer, say $k\, \Tr_{N-1}
F'^{3}$ if normalized to the fundamental. It is then clear that one
can extrapolate this to obtain the full $\mathcal H_\theta$ anomaly,
which is simply $k\, \Tr_N F^3$. On the other hand, if $N=3$ there is
no irreducible anomaly for the $SU(2)$ subgroup and one would have to
track the reducible (mixed) gauge anomalies instead in order to
reconstruct the irreducible $SU(3)$ anomaly. However since there are
only a few instances where this occurs we simply have checked case by
case that all these anomalies are integer as well.\footnote{ In many
cases it is possible to embed $\mathcal H$ in different ways into
$\mathcal G$ so that $\mathcal H_\theta$ actually corresponds to
different factors of $\mathcal H$. For instance, consider the breaking
of $E_7\to SU(3)\otimes SU(6)$. Depending on the choice of the shift
vector, both $\mathcal H_\theta=SU(3)$ and $\mathcal H_\theta=SU(6)$
are possible. In the former case we can conclude that the $SU(6)$
anomaly is integer while in the latter case this holds for the $SU(3)$
factor. As both embeddings should be physically equivalent, we
conclude that no fractional coefficients occur.}  This concludes our
proof that all the irreducible brane anomalies are integer.

To conclude this section we would like to comment on the relation
between the local 4d spectrum created by the bulk fermions and the
localized anomaly. With local 4d spectrum we mean the fields that are
not forced to vanish at the singularity (note that we do not include
brane fermions in this definition). Naively, one could think that this
spectrum determines the anomaly appearing at the fixed point.  This is
true only for $n=2,3$ where the anomaly computed from the local
spectrum equals four ($n=2$) or three ($n=3$) times the localized
anomaly.\footnote{This result can serve as an illustration that the
localized anomaly is not integer before the bulk is made anomaly
free.} In these cases one can embed the singularity in a torus
orbifold with four or three equivalent fixed points (see
Sec.~\ref{compact}). The zero mode spectrum is the same as the local
spectrum and the anomaly of the former is distributed evenly over the
fixed points.  For $n>3$ the localized anomaly appears to be rather
unrelated to the local spectrum.


\subsection{Localized Gravitational Anomalies in 6d Orbifolds}
\label{grav}

In this subsection we are interested in possible anomalies of general
coordinate transformations (sometimes referred to as Einstein
transformations, denoted by $\delta_{GC}$) and local Lorentz
transformations ($\delta_L$). Pure gravitational anomalies are known
to exist in $d=2 \mod 4$ dimensions, while mixed gauge-gravitational
ones can appear in any dimension. Moreover, it has been shown
\cite{Bardeen:1984pm} that anomalies of local Lorentz and general
coordinate invariance are equivalent in the sense that one can
introduce counterterms (local functionals of the metric/vielbein) which
violate both symmetries and can be used to shift the anomaly to either
pure Lorentz or pure Einstein form \cite{Bardeen:1984pm}. In other
words, these counterterms can be used to cancel one kind of anomaly
but not both. An Einstein anomaly leads to non-conservation of
energy-momentum while a Lorentz anomaly results in an energy-momentum
tensor that is not symmetric.  We will compute both type of anomalies
and find that the anomaly naturally occurs in Lorentz form.

We will adopt the following index convention: Greek letters take
values $0\dots 3$, lower case Latin letters $4,5$ and capital Latin
letters take values $0..5$. Moreover, letters from the beginning of
the alphabet ($A,B,a,b,\alpha,\beta$) are flat space (Lorentz)
indices, while letters from the middle of the alphabet
($M,N,m,n,\mu,\nu$) denote curved (Einstein) indices.  Notice that at
the singularity we can freely split vielbein and metric into purely
four and extra dimensional components as the mixed-index quantities
such as ${e_\mu}^a$ are zero by virtue of the boundary conditions (or,
equivalently, the orbifold projection). However, care is needed if
extra dimensional derivatives of these quantities occur.
 
In particular, we are interested in those transformations generated by
$\xi^m$ and $\Lambda_{ab}$,\ i.e.~the ``extra dimensional
transformations''. In the following we will refer to them as remnant
gravitational symmetries. From the point of view of an observer living
at the singularity they are merely gauge symmetries with the
important peculiarity that the corresponding connections are actually
composite fields made out of the vielbein ${e_m}^a$.  Although these
symmetries are eventually broken spontaneously in the flat vacuum and
only a global $SO(4)\otimes SO(2)$ symmetry survives, we need to make
sure that they are anomaly free in a generic gauge and gravitational
background to ensure quantum conservation of
energy-momentum.\footnote{Anomalies of these symmetries have been
studied in the context of $D$-branes where they are also known as
``normal-bundle anomalies''~\cite{Witten:1996hc}.}

Consistency with the orbifold projection implies that $\Lambda_{ab}$
is non-vanishing at the fixed point (its antisymmetry implies that its
parity is given by $\det P_L=1$). On the other hand $\xi^m$ has to
vanish at the fixed point, while its derivatives may be unconstrained:
For generic $Z_n$ orbifolds $\partial^{[n}\xi^{m]}$ and
$\partial_m\xi^m$ are non-vanishing while for the special case $n=2$
any first order derivative $\partial_{n}\xi^{m}$ survives the
projection.

First consider general coordinate invariance $\delta_{GC}$. The
fermion transforms simply as
\be
\delta_{GC} \psi = \xi^m\partial_m \xi\,.
\ee
We could have added a compensating local Lorentz transformation to
render the derivative covariant, but these additional terms would
cancel due to the boundary condition. By replacing the gauge
transformation $\Lambda$ with the operator $\xi^m\partial_m$ in
Eq.~(\ref{master}) and evaluating the trace with the heat kernel
techniques developed in App.~\ref{heatkernel}, we find for the anomaly
\be
\mathcal A_{GC} = - \frac{1}{(2 \pi)^3} 
\partial_m \xi^m \Tr\left( A_1 +  A_2 \right) F_0 \,
\delta(y)\,.
\ee
where we have defined the quantity
\be
A_{1}
=\frac{1}{96}\epsilon^{\alpha\beta\gamma\delta}
 {R_{\alpha\beta mn}}{R_{\gamma\delta}}^{mn}\,,
\label{A1}
\ee
and $A_2$ has been defined in Eq.~(\ref{A2text}).
We can see that only the ``scale transformations'' generated by
$\partial_m \xi^m$ are anomalous.\footnote{This is quite unlike the
case of Einstein anomalies on smooth manifolds, where at most the
$SO(d)\subset GL(d)$ subgroup generated by the antisymmetric part of
the tensor $\partial_m\xi^n$ is anomalous.} The latter constitute a
(non-compact) $U(1)$ subgroup of the $GL(2)$ transformations generated
by $\partial_m\xi^n$, and hence it is already in
its consistent form.  As a matter of fact, we can get rid of this
anomaly by adding the following local non-polynomial counterterm to the
action
\be
S_{CT} = \frac{1}{(2 \pi)^3} h \Tr F_0 (A_1+A_2) \delta(y)\,,
\label{GCCT}
\ee 
where $h\equiv \log\det {e_m}^a$ acts as a Goldstone boson for scale
transformations:
\be
\delta_{GC} h = \partial_m\xi^m\,.
\ee
The counterterm $S_{CT} $ has an anomalous variation that exactly
cancels the anomaly $\mathcal A_{GC}$. It is very similar to the kind
of counterterms found in Ref.~\cite{Bardeen:1984pm} that shift the
anomaly back and forth from Lorentz to Einstein form.  Note that the
above counterterm is in fact invariant under $\delta_L$ but does
generate an additional contribution to the trace anomaly. However, as
the conformal symmetry is violated anyways (at least by quantum
effects), we are not concerned by this fact.

To compute the Lorentz anomaly, we use for the generator
$\Lambda=\Lambda_{45} J^{45}$, where $J=- \sigma^3/2$
(see App.~\ref{conv}) and $\Lambda_{45}(x)$ is the local angle of
rotation. We thus have to evaluate:
\be
\mathcal A_L= \frac{1}{(4\pi)^{2} n}
\Lambda_{45}
\sum_{k=1}^{n-1} 
\frac{\cos(\pi k/n)}{ \sin^2(\pi k/n)}
  \Tr\, (P_{\mathcal G})^k A_2\, \delta(y)\,.
\ee
The difference to the gauge anomaly lies in the trace over the
$\Gamma$ matrices: The additional factor of $\sigma_3$ cancels the one
coming from the chirality matrix $\Gamma$, giving rise to a factor of
$\cos(\pi k/n )$ instead of $\sin(\pi k/n)$. Notice that this implies
that for $n=2$ there is no Lorentz anomaly.  Using relation
Eq.~(\ref{sum3}) we cast the Lorentz anomaly into the form
\be
\mathcal A_L= \frac{1}{8 \pi^2}\Lambda_{45}
\Tr
\left[
\left\{
n \left(\frac{F_0}{2\pi }\right)^2-\frac{1}{12}\left( \frac{R_0}{2\pi}\right)
\right\} A_2
\right]
\, \delta(y)\,,
\label{lor}
\ee
where we have defined the constant\footnote{It can be shown that this
is the amount of scalar curvature localized at a conical singularity
of opening angle $2\pi/n$ for arbitrary real $n$. }
\be
R_0 = 2\pi \left(n-\frac{1}{n}\right)\,. 
\ee
The occurrence of a local Lorentz anomaly signals a quantum energy
momentum tensor that is not symmetric. In particular, in our example
there is a contribution to $T_{45}-T_{54}$ that is proportional to a
$\delta$-function. One could easily write down the counterterms of
Ref.~\cite{Bardeen:1984pm} to shift this anomaly to Einstein form. We
would find non-conservation of energy momentum with a source term that
is proportional to a derivative of a $\delta$-function. Let us try to
interpret the form of the Lorentz anomaly, Eq.~(\ref{lor}).  Call the
remant $SO(2)$ local Lorentz symmetry $U(1)_L$.  Since it is an
Abelian anomaly, conversion into consistent form only requires
dividing the pure $U(1)_L^3$ anomaly by $3$.  Defining the $U(1)_L$
field strength
\be
R'=\frac{1}{2}{R_{\mu\nu}}^{45}dx^\mu dx^\nu
\ee
one can write the consistent anomaly as
\begin{multline}
\mathcal A^{\rm cons}_L= \frac{1}{8 \pi^2}\Lambda_{45}
\Tr
\left[
\biggl\{
n \left(\frac{F_0}{2\pi }\right)^2-\frac{1}{12}\left( \frac{R_0}{2\pi}\right)
\biggr\}\right.\\
\left.\biggl\{
\frac{1}{24} R_{SO(4)}^2+\frac{1}{36} R'^2 -\frac{1}{2}F^2
\biggr\}
\right]
\, \delta(y)\,,
\label{lorconsistent}
\end{multline}
The trace in Eq.~(\ref{lorconsistent}) goes over $\mathcal G$ and
$SO(4)$ indices.  Notice that $F_0^2$ is a nontrivial $\mathcal G$
valued matrix while $R_0$ is just a constant. As before, $F_0$
commutes with all $\mathcal H$ generators and hence it is also a
constant on each irreducible $\mathcal H$-representation.  Notice that
as far as the bulk anomalies are concerned, the second and third terms
in Eq.~(\ref{poly6}) (that is the reducible anomalies) are usually not
canceled amongst 6d fermions alone but require the existence of
Green-Schwarz two-forms with appropriate anomalous couplings. The
natural question is thus whether the anomaly Eq.~(\ref{lorconsistent})
can be canceled by adding suitable brane localized counterterms to
this Green-Schwarz Lagrangian.  As explained after Eq.~(\ref{inflow})
this requires the anomaly to be a product of a bulk and a brane form.
Clearly, this is not the case for the anomaly
Eq.~(\ref{lorconsistent}).  While terms proportional to
\be
\Tr F^2\,,\qquad\Tr R^2=\Tr R_{SO(4)}^2+2R'^2
\ee
could be canceled that way, terms proportional to
\be
\Tr F_0^2 F^2
\ee
and others remain.
One concludes that localized fermions or axions are needed for their 
cancellation. This may come as a surprise, as localized matter is not
necessarily expected to be charged under the $U(1)_L$ symmetry from a
purely field theory point of view, although it is a logical
possibility (i.e.~such couplings are allowed by the unbroken
gravitational symmetries).  On the other hand, if the field theory is
the low energy limit of some string theory, localized
fields originate from twisted string states, i.e.~closed strings on
the orbifold that encircle the conical singularity. They inherit the
transformation under any unbroken gravitational symmetry just as they
do for conventional gauge symmetries.  It is therefore quite
interesting that purely field theoretical anomaly cancellation
arguments predict that localized states are charged under the bulk
gravitational symmetries and hence they could naturally be zero modes
of twisted closed strings.

Let us conclude this section by noting that for a complete treatment
of the subject we would have to include the contribution of other
chiral bulk fields of different spin such as gravitinos or self dual
forms. This is outside the scope of the present paper and is left to
future research.

\section{Compactifications}
\label{compact}

Up to now we have dealt with orbifolds of infinite volume, i.e.~4d
Minkowski space times the infinite cone obtained by modding out
$\mathbb R^2$ by a discrete $Z_n$ subgroup of its $SO(2)$
symmetry. Naturally, the phenomenologically interesting models are
based on compact spaces such as the torus $T^2$. However, the vicinity
of the fixed points of such orbifolds is totally indistinguishable
from the infinite cone and the compactness of the space does not at
all influence the localized anomaly occurring there.  This section
serves to describe in more detail the embedding of the conical
singularities into compact orbifolds.

\begin{figure}[htb]
\includegraphics[width=\linewidth]{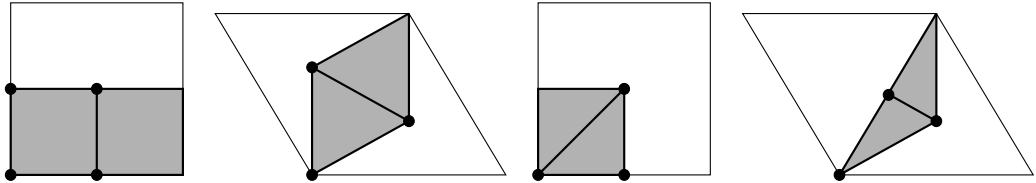}
\caption{\em The four different $Z_n$ orbifold geometries in 6d,
corresponding to $n=2,3,4,6$ (from left to right). We show the
embedding of the orbifold fundamental domain (shaded) in the torus
(thin line) as well as the fixed points (dots). The shaded regions
have to be folded over the center line and the edges (thick lines)
have to be identified. The resulting geometries are ``pillows'' with
three or four corners. Note that the edges correspond to nonsingular
bulk points.}
\label{orbs}
\end{figure}

The kind of orbifold singularities considered so far occur most
commonly in models that are obtained by modding out a two dimensional
torus $T^2$ by a discrete finite symmetry. It is well known that the
only subgroups of $GL(2)$ that map the torus lattice onto itself are
the $Z_n$ rotations with $n=2,3,4,6$.  The topology of the resulting
spaces are ``pillows'' where the two sides of the pillows represent
the bulk and the corners the conical singularities or fixed points. We
depict the four possibilities in Fig.~\ref{orbs}. Notice that $T^2/Z_4$ contains two $Z_4$ and one $Z_2$ singularity and  $T^2/Z_6$ contains one $Z_2$, $Z_3$ and $Z_6$ singularity each.
The $Z_n$ rotation has to be represented on the fields\footnote{For
the sake of simplicity we only treat the case of scalars in this
section. The generalizations to particles of nonzero spin involves the
matrix $P_L$ in addition to $P_{\mathcal G}$ and is straightforward.}
and defines a Wilson line $P_{\mathcal G}$ around the fixed point
located at the origin. In addition, one can allow for continuous or
discrete Wilson lines associated to lattice
translations~\cite{Ibanez:1986tp}. However, geometric constraints
apply. For instance, any lattice translations $T^\lambda$ (along some
lattice vector $\lambda$) and discrete $Z_n$ rotation $P$ satisfy
\be PT^\lambda P^{-1}=T^{P\lambda}\,,\qquad T^\lambda T^{\lambda'}=
T^{\lambda'}T^\lambda=T^{\lambda+\lambda'}\,.
\label{cons}
\ee
(Note that $P\lambda$ is another lattice vector).  Both relations
follow from elementary geometry in two dimensions.  The representation
of $T^{\lambda}$ on the fields (the torus Wilson line
$T^{\lambda}_{\mathcal G}$) and the $Z_n$ Wilson line $P_\mathcal G$
around the origin then have to obey analogous constraints. As a matter
of fact, if the torus Wilson lines are of finite order,
$(T^{\lambda}_\mathcal G)^m=1$, one can interpret the geometry as an
orbifold of a larger torus with no Wilson lines on them
\cite{vonGersdorff:2003dt}. The resulting orbifold group can be
nonabelian, which then allows for rank reducing breakings of the gauge
group \cite{Hebecker:2003jt,Ibanez:1987xa}.  However, a simple
solution to the constraints Eq.~(\ref{cons}) consists of taking
$P_{\mathcal G}$ to be an inner automorphism and choosing
$T^{\lambda}_\mathcal G$ to lie in the Cartan torus as well. In that
case we have\footnote{To prove this one has to use the geometric
identity $\sum_{k=0}^{n-1} P^k\lambda=0$.}  $(T^{\lambda}_\mathcal
G)^n=1$ and the resulting orbifold group is Abelian.

\begin{figure}[htb]
\includegraphics[width=\linewidth]{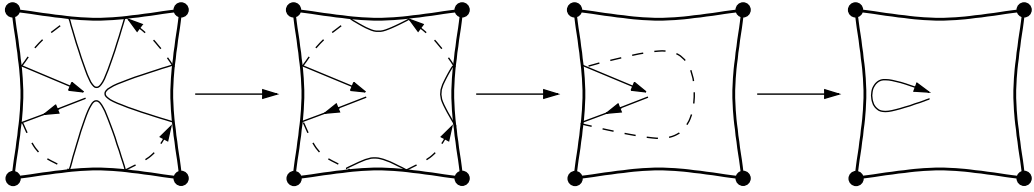}
\caption{\em Illustration of the second relation in Eq.~(\ref{cyclic})
for the $Z_2$ pillow. If the path encloses the four corners of the
space in a cyclic order, it becomes contractible and the associated
Wilson line must be unity (the dashed lines represent parts of the
path on the back side of the pillow).}
\label{contract}
\end{figure}

We will refer to this description in terms of the covering torus and
its Wilson lines as the {\em upstairs} picture. On the other hand, the
maybe simpler and more physical approach is the {\em downstairs}
picture in terms of the fundamental domain of the orbifold, i.e.~the
pillow geometries depicted in Fig.~\ref{orbs}. To each of the corner
singularities of these spaces one associates a Wilson line
$P^{(f)}_{\mathcal G}$ corresponding to the rotation around that given
fixed point which is a product of torus and $Z_n$ Wilson lines.  The
geometric constraints of the upstairs picture then translate into
constraints onto the $P^{(f)}_{\mathcal G}$, for instance
\be
\bigl(P^{(f)}_{\mathcal G}\bigr)^{n_f}=1\,,\qquad
P^{(1)}_{\mathcal G}
P^{(2)}_{\mathcal G}\cdots  P^{(\nu)}_{\mathcal G} =1\,,
\label{cyclic}
\ee
where $\nu$ is the number of fixed points and $n_f$ is the order of
the fixed point. The ordering of the Wilson lines is counterclockwise
along the corners of the pillow and cyclic permutations of the last
identity hold.
While the meaning of the first relation in Eq.~(\ref{cyclic}) is
obvious, the second identity has another simple geometric
interpretation that is illustrated in Fig.~\ref{contract}.
If the $T^{\lambda}_{\mathcal G}$ commute with $P_{\mathcal G}$ the
only additional constraints to Eq.~(\ref{cyclic}) are that the
$P_{\mathcal G}^{(i)}$ commute. For nonabelian orbifolds however, the
downstairs constraints become more complicated. In the most general
case, the $P_{\mathcal G}^{(i)}$ have to form a representation of
the (nonabelian) fundamental group of the pillow space.
We refer the reader to Ref.~\cite{Hebecker:2003jt} for some possible
generalizations of this downstairs construction that cannot be lifted
to the upstairs picture.

In the downstairs picture it is particularly transparent that the
compact torus orbifolds are locally equivalent to non-compact orbifolds
and all one has to know for the physics of the fixed points are the
local Wilson line one encounters by encircling the singularities. In
particular, the localized anomaly is completely known from this local
data.

We would like to perform a consistency check on the local Lorentz
anomaly found in Sec.~\ref{grav} by taking the
compactification limit. To this end, consider first a torus
compactification and take the limit of vanishing internal volume.
Clearly the remnant local Lorentz symmetry is spontaneously broken,
but in the limit where all heavy KK modes decouple it survives as a
global symmetry. As can easily be checked, on the 4d zero modes of a 6d
Weyl fermion this symmetry acts as a chiral $U(1)$:
\be
\left(\begin{array}{c}\psi_L\\\psi_R\end{array}\right)\to
\exp\left(-\frac{i}{2}\Lambda_{45}\gamma^5\right)
\left(\begin{array}{c}\psi_L\\\psi_R\end{array}\right)\,.
\ee
Hence the spectrum of zero modes is not anomaly free but instead we
find\footnote{ Of course, this anomaly can also be obtained from the
expression for the 6d anomaly. The reason why this term appears to be
absent for finite $R$ is that the limits $s\to 0$ ($s$ is the
regulator used to compute the anomaly) and $R\to 0$ ($R$ being the
volume modulus of the torus) do not commute. If one takes the latter
limit first one indeed arives at Eq.~(\ref{torus}).
}
\be
\mathcal A_{L,tor}^{(0)}=2\cdot\frac{1}{8\pi^2}\Lambda_{45} \Tr A_2\,.
\label{torus}
\ee
Note that this is unlike the case of gauge symmetries which always
result in vector-like theories in 4d.

Now orbifold this model on $T^2/Z_n$.  The Wilson lines around the
singularities determine which of the two zero modes (if any) survives
the projection.  Let us -- for simplicity -- assume a $U(1)$ gauge
symmetry.  The expected zero mode anomaly is
\be
\mathcal A_{L,orb}^{(0)}=
n_0\cdot \frac{1}{8 \pi^2} \Lambda_{45} 
\Tr  A_2  \,,
\ee
where the integer coefficient $n_0$ counts the number of zero modes
(only $n_0=0,1$ occur). Note that the anomaly is the same independent
of the chirality of the 4d zero mode.  The localized local Lorentz
anomaly, integrated over the compact dimension, should equal the
difference of the zero mode anomaly for the chiral $U(1)$ between the
orbifold and torus compactification:
\be
\int \mathcal A_L   
=\mathcal A_{L,orb}^{(0)}-\frac{1}{n}\mathcal A_{L,tor}^{(0)}\,.
\label{difftororb}
\ee
Eq.~(\ref{difftororb}) can be checked from Eq.~(\ref{lor}) directly
for $n=2,3,4,6$ and for arbitrary $P_{\mathcal G}$. In particular, for
$n=2$ there is always precisely one zero mode, independent of the
Wilson line, implying that Eq.~(\ref{difftororb}) gives zero in
accordance with our analysis of Sec.~\ref{grav}.

\section{Examples in Supersymmetric Theories}
\label{examples}

In this subsection we would like to illustrate our so far very
abstract discussion with some examples that are possibly relevant to
6d supersymmetric orbifold GUTs. We will focus mainly on the
irreducible gauge anomaly of type $\Tr F^3$ as it gives the most
stringent restrictions on the localized spectrum.  After commenting on
the SUSY breaking features of a 6d orbifold fixed point, we will
review 6d bulk anomaly cancellation conditions and finally apply the
results of Sec.~\ref{anomalies} to some interesting GUT gauge groups.

\subsection{Supersymmetry breaking}
\label{susybreaking}

The main success of orbifolds is their capacity to produce chiral
boundary-conditions and spectra. Given that the orbifold breaks
chirality it is not surprising that SUSY is also broken at an orbifold
fixed point.  Due to the presence of the $P_L$ factor in the orbifold
twist particles acquire spin-dependent boundary conditions.  Moreover,
in general there is an additional explicit $R$-symmetry twist due to
the fact that $(P_L)^n=-1$ for fermions. For instance, while for gauge
fields we have $(P_{\mathcal G})^n=+1$, for gauginos we actually need
$(P_{\mathcal G})^n=-1$. One concludes that $P_\mathcal G$ has to
include an additional-$R$ symmetry twist. How much SUSY is locally
broken depends on the order $n$ of the fixed point as well as the form
of the $R$-symmetry twist, as we will see shortly. The amount of
global (low energy) SUSY breaking depends on the (mis)match of the
supercharges locally preserved at {\em different} fixed points. The
latter is generally referred to as Scherk-Schwarz (SS) SUSY breaking
and is widely used in string and field theoretical models. To find the
surviving supercharges at a given singularity, we have to look more
closely at the $R$-symmetry. Pure $N=(1,0)$ supersymmetric gauge
theory in 6d is invariant under an $SU(2)_R$ symmetry, which acts on
gauginos only but not on the gauge fields. Gauginos are doublets under
this symmetry and obey a symplectic Majorana constraint,
\be
(\psi^i)^*=\epsilon_{ij} C \psi^j\,,
\label{SM}
\ee
where $C$ is the 6d charge conjugation matrix and $\epsilon_{ij}$ is
the antisymmetric tensor of $SU(2)_R$. Without loss of generality we
can choose the $R$ symmetry twist to be\footnote{This might not
simultaneously be possible at different fixed points, in which case we
encounter SS supersymmetry breaking. See Ref.~\cite{Braun:2006se} for
a recent application in six dimensions.} $\exp(2 \pi i\, p_R/n\,
\sigma_3)$ with $p_R=r-1/2$. Eliminating $\psi^2$ with the help of the
constraint Eq.~(\ref{SM}), the remaining $\psi^1$ is simply twisted by
a phase\footnote{For $n$ odd and the Lorentz twist $\tilde P_L$ (See
Sec.~\ref{anomalies}) the additional $R$-symmetry twist is integer,
$p_R=r$. We will not present any examples for this case which in fact
works very similar. }
\be
P_R=\exp(2\pi i \, p_R/n )\,.
\label{Rtwist}
\ee
In terms of the 4d chiral spinors $\psi_L$ and $\psi_R$ (see
App.~\ref{conv}) the orbifold constraints on the gauginos of $\mathcal
H$ imply\footnote{Flipping the 6d chirality of the gaugino just flips
the 4d chirality in Eq.~(\ref{gauginofp}). }
\begin{eqnarray}
\psi_L(y=0)&=&\exp(2\pi i\, (r-1)/n )\,\psi_L(y=0)\,,\nonumber\\
\psi_R(y=0)&=&\exp(2 \pi i\, r /n )\,\psi_R(y=0)\,.
\label{gauginofp}
\end{eqnarray}
We see that for $r=0\mod n\ (r=1 \mod n)$ it is $\psi_R\ (\psi_L)$
that locally survives the projection, leading to $N=1$ unbroken SUSY.
For any other value of $r$ both 4d gauginos vanish, implying complete
breakdown of SUSY.  For later purposes let us introduce the periodic
Kronecker symbol $\delta_{r,n}$ which by definition equals one for
$r=0\mod n$ and zero otherwise. The amount of 4d supersymmetry at the
fixed point can then be expressed as
\be
N=\delta_{r,n}+\delta_{r-1,n}\,.
\ee
One sees that $n=2$ fixed points always have $N=1$ supersymmetry,
while for $n>2$ one can have $N=1$ or $N=0$. Note that in the $Z_4$
and $Z_6$ compact orbifolds there is actually a $Z_2$ fixed point
present. This means that locally some SUSY survives in these cases.

Let us now introduce matter hypermultiplets. The $R$-symmetry twist
acts on the hyperscalars only, implying that we now need $(P_{\mathcal
G})^n=-1$ for the multiplet as a whole. In general, this requires
additional Wilson lines for matter fields.  We will see some examples
in Sec.~\ref{GUTexamples}.  It is important to check that any symmetry
used in orbifolding is anomaly free in the unorbifolded theory
\cite{GrootNibbelink:2002qp}. For the $SU(2)_R$ twist(s) this is
trivially the case. If the additional Wilson lines in the matter
sector involve $U(1)$'s, this may provide further constraints.

Next, we would like to comment on models with extended
$N=(N_+,N_-)$ SUSY. The $R$-symmetry group is $U\!Sp(2N_+)\otimes
U\!Sp(2N_-)$~\cite{deWit:1997sz}. In particular, for the case of
$N=(1,1)$ SUSY, it is $SU(2)_R\otimes SU(2)_R'$. The (say) left handed
vector multiplet is paired up with a right handed adjoint
hypermultiplet to form a complete $N=(1,1)$ vector multiplet. With
respect to the second $SU(2)_R'$, the vector multiplet is a singlet
and the hypermultiplet a doublet obeying the reality constraints
\be
(H^a_i)^*=\epsilon_{ab}\epsilon^{ij} H_j^b\,\qquad 
(\psi^a)^*=\epsilon_{ab} C \psi^b\,. 
\ee
Without loss of generality we can choose the $SU(2)_R'$ twist to be
proportional to $\sigma_3$ and eliminate the $a=2$ components of the
hypermultiplet with the above constraint. The action of the total $R$
symmetry twist on $\psi^1$ can then be parametrized as
\be
P_R'=\exp(2 \pi i p_R'/n)\,,\qquad p_R'=r'-\frac{1}{2}\,.
\ee
The amount of supersymmetry surviving at the fixed points can again be
computed by determining the non-vanishing fields at the singularity. It
turns out to be
\be
N=\delta_{r,n}+\delta_{r',n}+\delta_{r-1,n}+\delta_{r'-1,n}\,.
\label{extended}
\ee
One sees that for $n=2$ only $N=2$ occurs, while for $n>2$ we can
break down to $N=2$, $N=1$ and $N=0$. This has lead people to consider
$n>2$ orbifolds \cite{Hall:2001xr}. However as far as low energy
supersymmetry is concerned one can achieve $N=1$ on a $Z_2$ orbifold
by invoking the SS mechanism.

\subsection{Bulk anomaly cancellation}

In this subsection we compute anomaly free spectra in 6 spacetime
dimensions. We will focus here on the irreducible gauge anomaly only,
that is the nonfactorizable trace of four generators.  In addition,
the cancellation of the irreducible gravitational anomaly has to be
enforced.  Besides spin $1/2$ fermion, there are contributions to the
irreducible gravitational anomaly from gravitinos and (anti) self-dual
tensor fields. Finally, the reducible anomalies, e.g.~$(\Tr F^2)^2$ or
$\Tr R^2\Tr F^2$, can be canceled by the GS-mechanism involving
two-forms,\cite{Green:1984sg} and do not place any restrictions on the
6d fermion spectrum.  For previous work on 6d anomaly cancellation see
Refs.~\cite{ Erler:1993zy, Asaka:2002my,
Hebecker:2001jb
}.

The simplest GUT group, $SU(5)$, does not lead to nontrivial chiral
6d spectra. The reason is that 6d matter hypermultiplets have to
carry opposite chirality w.r.t.~the gaugino (See, e.g., the discussion in
\cite{Hebecker:2004xx}). Furthermore, complex conjugation does not
flip the chirality of a 6d spinor, in contrast to the 4d case. In
accordance with this $\Tr F^4$ is the same for a representation and
its conjugate. Cancellation of the irreducible part of the gauge
anomaly requires the following identities for $SU(5)$:
\begin{eqnarray}
\Tr_{\bf 24} F^4 &=& 10\Tr_{\bf 5} F^4+6(\Tr_{\bf 5}F^2)^2\,,\nonumber\\
\Tr_{\bf 10} F^4 &=& -3\Tr_{\bf 5} F^4+3 (\Tr_{\bf 5}F^2)^2\,,\nonumber\\
\Tr_{\bf 15} F^4 &=& 13\Tr_{\bf 5} F^4+3 (\Tr_{\bf 5}F^2)^2\,,
\label{su5identity}
\end{eqnarray}
We thus conclude that with fundamental and antisymmetric
representations alone no ``economic'' anomaly free models are
possible: 6d anomaly cancellation requires in fact no less than 10
fundamentals. The simplest nontrivial solutions would be an
antisymmetric ($\rep {10}$) and a symmetric ($\rep{15}$) or an adjoint
multiplet of opposite chirality. The latter of course leads to a
vector-like theory in 6d.  As for the former solution note that the
quarks contained in the symmetric and antisymmetric representations
and the ones in the adjoint have incompatible hypercharges.  This
means that one has to project out one of them by orbifolding.

In fact these two solutions generalize to $SU(N)$,\footnote{See
Ref~\cite{Hall:2001zb,Hebecker:2003we} for 6d models with an $SU(6)$
GUT group.} as the $SU(5)$ relations Eq.~(\ref{su5identity}) have a
straightforward generalization
\be
\Tr_\rep{r} F^4=c_4\Tr_N F^4 + c_2\Tr_N (F^2)^2\,,
\ee  
with $c_4=2N,N+8,N-8$ and $c_2=6,3,3$ for the adjoint, symmetric
tensor and antisymmetric tensor representations respectively.

In the $SO(10)$ case there do exist interesting anomaly free 6d
spectra as can be seen from the relations
\begin{eqnarray}
\Tr_{\rep{45}} F^4 &=& 2\Tr_{\bf 10} F^4+3(\Tr_{\bf 10}F^2)^2\,,\label{45}\\
\Tr_{\rep{16}} F^4 &=& -\Tr_{\bf 10} F^4+\frac{3}{4}(\Tr_{\bf 10}F^2)^2\,,\label{16}
\end{eqnarray}
The minimal matter content would thus consist of two $\rep{10}$
hypermultiplets in the bulk, while further $\rep {10}$'s and
$\rep{16}$'s should occur in pairs.  Finally, $E_6$, $E_7$ and $E_8$
do not have any independent fourth order invariants and thus the
adjoint representation (as any other representation) does not have an
irreducible bulk anomaly.

\subsection{Localized anomaly cancellation}
\label{GUTexamples}

Since the fixed points are four-dimensional, only $SU(N)$ factors of
$\mathcal H$ will potentially contribute to the irreducible gauge
anomaly. We will convert all anomalies to the fundamental, i.e.~we will
express everything in terms of the quantities
\be
I_{N}\equiv \Tr_{N}F^3\,.
\ee
For $\mathcal G=SU(5)$, let us fix the matter content to an adjoint
hypermultiplet of opposite chirality w.r.t.~the gauge multiplet.  The
local breaking to the standard model is achieved by deleting the third
(equivalently second) node of the Dynkin diagram, which is accomplished
by the shift vector
\be
V=\frac{1}{n}\,\mu^{(3)}\,.
\label{su5}
\ee
The $\rep{24}$ branching rule is given by
\be
\rep{24} \to 
(\rep{3},\rep{2})_{1}\oplus(\crep{3},\rep{2})_{-1}
\oplus \rep{adj}({\rm SM})\,,
\ee
where the generator of the $U(1)$ is given by $n V=\mu^{(3)}$. As
explained in the Sec.~\ref{susybreaking}, we need to multiply this
gauge twist by an $R$-symmetry twist Eq.~(\ref{Rtwist}). For the
matter hypermultiplet an additional phase $\exp(2\pi i p_M/n)$ with
$p_M=m-1/2$ has to be added. This can be embedded in a matter $U(1)$,
which is non anomalous in the 6d unorbifolded theory as the trace over
three adjoint generators vanishes.  However, since this system is
actually $N=(1,1)$ supersymmetric, we can also interpret $p_M$ as the
other $R$ symmetry twist $p_R'$ (see Sec.~\ref{susybreaking}) and
identify $m$ with $r'$. Following Sec.~\ref{localized} one finds that
the gaugino contributes
\begin{multline}
\tilde I_{\rep{24}}
=\left(
-\frac{1}{n}-\frac{p_R}{n}
+\floor{1/n+p_R/n}
\right)I_{\rep{3}}
\\
+\left(
\frac{1}{n}-\frac{p_R}{n}
+\floor{-1/n+p_R/n}\right)I_\crep{3}
\end{multline}
As before, $\floor{x}$ stands for the floor function defined as the
closest integer $k\leq x$.  The two floor functions cancel unless
$r=0,1\mod n$ (in which case they give unity), hence
\be
\tilde I_{\rep{24}}=2\left(-\frac{2}{n}+
\delta_{r,n}+\delta_{r-1,n}
\right)I_{\rep 3}\,,
\ee
where the Kronecker-$\delta$ has again to be continued periodically.
The matter multiplet contributes $-\tilde I_\rep{24}$ with $r$
replaced by $r'$, hence
\be
\tilde I_\rep{24}-\tilde I_\rep{24'}=2\left(
\delta_{r,n}+\delta_{r-1,n}-\delta_{r',n}-\delta_{r'-1,n}
\right)I_{\rep 3}\,.
\ee
The fractional parts cancel as expected and the remaining $SU(3)$
anomaly has an integer coefficient.  Note that the inflow terms are
zero since the adjoint is a real representation.  We can see that if
the anomaly is non-vanishing it carries a coefficient 2, so that the
localized matter content has to be either one quark doublet $(\rep
3,\rep 2)$ or two quark singlets $(\rep3,\rep1)$. Which of the quarks
doublets contained in the $\rep{24}$ have zero modes upon
compactification depends on the Wilson lines at the other fixed
points. An interesting model based on $T^2/Z_3$ can be constructed as
follows: Break $SU(5)$ to the standard model with the gauge twist
Eq.~(\ref{su5}) and supersymmetry to $N=1$ with the $R$-symmetry twist
$r=0$ and $r'=2$ at all three fixed points. This produces three quark
doublets $(\rep3,\rep 2)_{1}$ from bulk zero modes, whose anomaly is
distributed uniformly over the three fixed points.  The model can then
trivially be made anomaly free by completing the generations at each
of the fixed points.  Of course, Higgs fields have to be added to the
model.  However, there does not exist a simple $SU(5)$ representation
that contains $SU(2)$ doublets with the correct hypercharge w.r.t.~to
the quark doublets, and, hence, the Higgs sector should reside on a
brane. On the other hand, this forbids some of the Yukawa couplings,
which, in addition, cannot be generated radiatively.  One probably has
to extend the gauge group to $SU(6)$ to include a realistic Higgs
sector in the model.  A similar model has been developed in
Ref.~\cite{Watari:2002tf} with $SO(10)$ as a gauge group and Higgs
fields at one of the fixed points. However apart from the problem of
how to generate Yukawa interactions for two generations, that model
requires an additional mechanism to break the remaining $SU(5)$ and in
particular also suffers from the same doublet triplet splitting
problem as conventional 4d GUT models.  Other -- less symmetric --
constructions are certainly possible, perhaps with the quark doublets
of only one or two generations originating from the bulk. Note that
there is no localized anomaly from the gauginos on fixed points with
unbroken $SU(5)$, hence one can always introduce complete generations
$(\rep{ 10},\crep 5)$ there.

Next, consider $\mathcal G=SO(10)$, which has been studied extensively
in the context of orbifold GUTs in
\cite{Asaka:2001eh,Hall:2001xr,Asaka:2002my,Watari:2002tf,Asaka:2003iy}.
Let us start with the breaking pattern $SO(10)\to SU(5)\otimes U(1)$
which can be achieved by deleting the fourth or fifth node of the
Dynkin diagram \cite{Hebecker:2003jt}, see Fig.\ref{dynkin}. It thus
corresponds to the shift vector
\be
V=\frac{1}{n}\,\mu^{(4)}\,.
\ee
The branching rules of the various low-dimensional representations
read as follows.
\begin{eqnarray}
\rep{45} &\to& \rep{1}_0 \oplus\rep{10}_1
\oplus\crep{10}_{-1} 
\oplus\rep{24}_0\,,\\
\rep{10} &\to& 
\rep{5}_{1/2}\oplus\crep{5}_{-1/2}\,,\\
\rep{16} &\to&
\rep 1_{-5/4}\oplus\crep{5}_{3/4}\oplus \rep{10}_{-1/4}\,.
\end{eqnarray}
Looking at the charges occurring in these branching rules, we see that
$n\,V_\rep{10}=1/2$ and $n\,V_\rep{16}=3/4$ (modulo
integers). One concludes that the $\rep{16}$ has to have an additional
orbifold phase $p_M=m-1/4$.  The latter can be introduced without
problems, since it can be embedded in an anomaly free\footnote{Any
$U(1)$ is trivially anomaly free in $d=6$ and $\mathcal G=SO(10)$, as
the trace of any three $SO(10)$ generators vanishes.} global $U(1)_M$
under which only the $\rep{16}$ is charged.  One can also include an
additional phase $p_M=m$ for the $\rep{10}$ in order to account for
further local or global Wilson lines.\footnote{In order to keep the
notation simple, we generically do not indicate the dependence of
$p_M$ or $m$ on the representation.}  Let us now focus on the possible
irreducible gauge anomaly at the 4d fixed point, that is the $SU(5)^3$
anomaly.  Normalizing to the fundamental of $SU(5)$ one finds
\be
\tilde I_{\rep {45}}
=\left(
-\frac{2}{n}
+\delta_{r,n}+\delta_{r-1,n}
\right)I_{\rep{5}}\,.
\ee
In a similar fashion, we find for the $\rep{10}$ and $\rep{16}$
\be
\tilde I_{\rep {10}}
=
\left(-\frac{1}{n}+\delta_{m,n}\right)I_{\rep{5}}\,,\qquad
\tilde I_{\rep{ 16}}=\tilde I_{\crep {16}}
=\left(\frac{1}{n}-\delta_{m,n}\right)I_{\rep{5}}\,.
\ee
Adding the contribution of a (say) left handed gaugino and two right
handed matter multiplets $\rep{10}$ and $\rep{10'}$, the bulk anomaly
is canceled.  As it must be, this fermion content also cancels the
non-integer coefficient of the $I_{\rep 5}$ as it is directly linked to
the 6d anomaly.  The total anomaly 
\be
\tilde I_{\rep{45}}-\tilde I_{\rep{10}}-\tilde I_{\rep{10'}}=
(\delta_{r,n}+\delta_{r-1,n}-\delta_{m,n}-\delta_{m',n})I_{\rep 5}\,.
\ee
again depends on the $R$-symmetry and matter phases. It vanishes for
certain combinations of $r$, $m$ and $m'$. For instance in the $n=2$
model of Ref.~\cite{Asaka:2001eh,Asaka:2002my} it was assumed that the
$\rep{10}$ had positive parity ($m=0$) and the $\rep{10'}$ negative
($m'=1$) and hence no anomaly occurs. In this case it turns out that
the local spectrum\footnote{Recall that we define the local spectrum
to be the nonzero bulk fields at the singularity.}  is anomaly free
for both $r=0,1$.  On the other hand consider an $n=4$ fixed point and
the assignments $r=0$, $m=0$ and $m'=2$. Still the localized anomaly
vanishes, but it is not hard to check that the local spectrum consists
of one $\rep{10}$ and two $\crep{5}$, which would not be anomaly
free. This example shows that one in general cannot determine the
localized anomaly from the knowledge of the local spectrum. Finally
notice that the localized anomaly of pairs of $(\rep{10},\rep{16})$ or
$(\rep{10},\crep{16})$ again has no fractional coefficient.

The orbifold twist
\be
V=\frac{1}{2}\mu^{(2)}
\ee
leaves the Pati Salam (PS) group $SU(4)\otimes
SU(2)\otimes SU(2)$  unbroken. The corresponding branching rules read
\begin{eqnarray}
\rep{45}&\to&(\rep 6,\rep 2,\rep 2)\oplus \rep{adj}(\rm PS)\,,\nn\\
\rep{10}&\to&(\rep 6,\rep1,\rep 1)\oplus (\rep 1,\rep 2,\rep 2)\,,\nn\\
\rep{16}&\to&(\rep 4,\rep 2,\rep 1)\oplus (\crep 4,\rep 1,\rep 2)
\end{eqnarray}
The localized anomaly has been discussed in the context of the model
of Ref.~\cite{Asaka:2002my}, we include it here for the sake of
completeness. As the $\rep 6$ is a real representation of $SU(4)$, it
does not have a 4d anomaly. Hence the only contribution can come from
the $\rep{16}$.  The phases for the $\rep 4$ and $\crep 4$ turn out to
be $V_{\rep 4}=1/2$ and $V_{\crep 4}=0$. To this, one has to add a
matter phase $p_M=m-1/2$.  From Eq.~(\ref{nonuniversal}) one finds for
the localized anomaly
\be
-\tilde I_{\rep{16}}
=\frac{1}{2 \pi i}\left(\log'[e^{2\pi i(\frac{1}{2}+\frac{m}{2}-\frac{1}{4})}]
-\log'[e^{2\pi i(\frac{m}{2}-\frac{1}{4})}]
\right)2 I_\rep{4}
=(-1,1)I_\rep{4}\,
\ee
where the two coefficients refer to $m=0,1$ respectively.  We observe
that unless further pairs of $(\rep{10},\rep{16})$ are introduced in
the bulk, one always needs to add at least a single $(\rep{4},\rep{1},\rep 1)$
or $(\crep{4},\rep{1},\rep 1)$ at the fixed point. This could possibly
correspond to a left handed quark in realistic orbifold
constructions.

Next we turn to $\mathcal G=E_6$. As mentioned before, $E_6$ does not
have an irreducible 6d gauge anomaly. Furthermore, like for $SO(10)$,
the trace over three generators vanishes so there cannot be any
anomaly inflow from the bulk to the fixed point.  Let us consider the
shift vector
\be
V=\frac{1}{3}\mu^{(3)}\,.
\ee
This eliminates the third node of the Dynkin diagram but projects in
the highest root $\theta$. The resulting gauge group is $SU(3)\otimes
SU(3)'\otimes SU(3)''$ where $\alpha^{(1,2)}$ serve as simple roots for
$SU(3)$, $\alpha^{(4,5)}$ for $SU(3)'$ and $\alpha^{(6)}$
and $-\theta$ for $SU(3)''$. The branching of the adjoint is given by
\begin{eqnarray}
\rep{78}&\to& (\crep{3},\rep{3},\rep{3})
\oplus (\rep3,\crep3,\crep3)\oplus\rep{adj}\nonumber
\\
&\to&(\crep{3},\rep{3},\rep{2})_{-1}\oplus(\crep{3},\rep{3},\rep{1})_{2}
\oplus (\rep3,\crep3,\rep2)_{1}\oplus (\rep3,\crep3,\rep 1)_{-2}
\oplus\dots\,,
\end{eqnarray}
where in the second line we have indicated the breaking of $SU(3)''$.
From the charges it is clear that $V_{\crep3\rep{33}}=2/3$
and $V_{\rep3\crep{33}}=1/3$ (modulo integers).   
One finds
\begin{eqnarray}
\tilde I_\rep{78}
&=&(3,3,-6)(I_\rep 3 -I'_\rep{3}-I''_\rep{3})\,,
\end{eqnarray}
where the coefficients refer to the values $r=(0,1,2)$.  For $r=0,1$
(corresponding to $N=1$ supersymmetry at the fixed point, see
Sec.~\ref{susybreaking}), the minimal localized matter content that
cancels the localized anomaly is given by $3\times
(\crep{3},\rep{1},\rep{1})$ and $1\times (\rep1,\rep{3},\rep{3})$, or
$3\times (\rep{1},\rep{3},\rep{1})$ and $1\times
(\crep3,\rep{1},\rep{3})$, or $3\times (\rep{1},\rep{1},\rep{3})$ and
$1\times (\crep3,\rep{3},\rep{1})$.  Let us identify the first $SU(3)$
with color and the second with $SU(3)_W \supset SU(2)_W$.  Ignoring
the $U(1)$ charges for a moment, one sees that e.g.~the first choice
provides three generations of leptons and quark singlets.  The missing
fields have to either reside on other fixed points or come from bulk
zero modes. For the latter possibility notice that the only bulk field
that is nonvanishing at the fixed point is a $(\rep
3,\crep{3},\crep{3})$.  Provided it is nonvanishing at the other fixed
points as well and, thus, has a zero mode, it precisely gives rise to
the missing quarks. In any event, one has to find a suitable $U(1)$
subgroup of $SU(3)_W\otimes SU(3)''$ that can act as hypercharge and
is in particular free of localized anomalies.  Building a concrete
orbifold model containing such a particular local GUT breaking at one
of its fixed points is beyond the scope of the present paper and is
left to future research.

As a last example consider the breaking of $E_7$ to
$SU(5)\otimes SU(3)\otimes U(1)$ accomplished by the shift vector
\be
V=\frac{1}{n}\mu^{(4)}\,,\qquad n>3\,.
\ee
(For $n=3$ the gauge group is actually enhanced to $SU(6)$).  The
adjoint of $E_7$ is the $\rep{ 133}$ and has the following branching
rule:
\be
\rep{133}\to
(\rep{10},\crep{3})_{1}\oplus(\rep{5},\crep{3})_{-2}\oplus
(\crep{10},\rep{3})_{-1}\oplus(\crep{5},\rep{3})_{2}\oplus
(\rep{5},\rep1)_{3}\oplus(\crep{5},\rep1)_{-3}\oplus \rep{adj}\,.
\ee
For instance, for $r=0,1$ (the $R$ symmetry twist that preserves $N=1$
supersymmetry) we find for the anomaly
\be
\tilde I_\rep{133}=I_\rep{5}-5 I_\rep{3}\,,
\ee
for any $n>3$. It turns out that at least three irreducible
representations are necessary to cancel this anomaly. The
possibilities are a $(\crep{5},\rep{3})$ and two $(\rep 5,\rep 1)$
(one or both of which can also be replaced by $(\rep{10},\rep 1)$) or
a $(\rep{10},\rep{3})$, a $(\crep{5},\crep{3})$ and a $(\rep 5,\rep
1)$. The latter exactly gives three generations of quarks and leptons
plus an additional $\rep 5$ which could serve as one of the Higgses.
Once again, including more matter fields in the bulk (in particular
extending bulk supersymmetry) changes the localized anomaly and hence
these predictions. For a model with $N=(1,1)$ supersymmetry and this
particular breaking pattern see Ref.~\cite{Hebecker:2003jt}.

We can see from the last two examples that if the unbroken subgroup
contains two or more nonabelian factors that have cubic anomalies in
4d, the localized anomaly cancellation condition can be quite
restrictive. Some (incomplete) list of breakings of the exceptional groups and their anomalies is presented in Tab.~\ref{exceptional}

\begin{table}[thb]
\begin{center}
\begin{tabular}{c|l|c}
Shift Vector\B &Breaking &Gaugino Anomaly $\tilde I_\rep{adj}$\\
\hline
\T$\mu^{(2)}/2$&$E_6\to SU(6)\otimes SU(2)$&0\\
\B$\mu^{(3)}/3$ & $E_6\to SU(3)\otimes SU(3)'\otimes SU(3)''$
&$3(I_\rep 3 -I'_\rep{3}-I''_\rep{3})$\\
\hline
\T
$\mu^{(2)}/3$
&$E_7\to SU(6)\otimes SU(3)$&$-5I_\rep{3}+2I_\rep 6$\\
$\mu^{(7)}/2$
\B&$E_7\to SU(8)$ &0\\
\hline
\T
$\mu^{(4)}/5$
&$E_8\to SU(5)\otimes SU(5)'$&$I_\rep{5}-7I'_{\rep5}$\\
$\mu^{(6)}/3$
&$E_8\to E_6\otimes SU(3)$&$9I_\rep{3}$\\
$\mu^{(8)}/3$
&$E_8\to SU(9)$&$-3 I_\rep{9}$\\
\end{tabular}
\caption{\it The localized anomaly for different breakings of the
exceptional groups $E_{6,7,8}$. Note the anomaly vanishes in 4d if
the adjoint branches exclusively into real representations of the
unbroken group. It is assumed that $N=1$ supersymmetry is preserved
$(r=0,1)$.}
\end{center}
\label{exceptional}
\end{table}

\section{Conclusions}

In this paper we have analyzed the occurrence of localized anomalies on
fixed points of orbifolds of 4+2 dimensional smooth spacetimes.  Such
fixed points constitute conical singularities obtained by gluing
together the edges of the ``wedge'' $\{r,\theta\}$, the angle theta
taking values $0\leq \theta\leq 2\pi/n$.  Internal (global or local)
symmetries can be broken by this procedure by identifying the 6d
fields on the two edges only up to an orbifold twist (Wilson line). 

We have shown that the pure gauge anomaly of the unbroken group
$\mathcal H\subset \mathcal G$ caused by a 6d Weyl fermion can be
interpreted as coming from a background field configuration that would
be obtained by gauging away the orbifold twist. This particular form
led us to a natural splitting of the anomaly localized at the 4d fixed
point into three terms
\be
-\Tr(V F^3)+\Tr(\floor V F^3)+\frac{1}{2}\Tr(F^3)
\ee
where $V$ is the so-called twist vector characterizing the orbifold
twist (an element of the Cartan torus of the algebra) and $\floor x$
denotes the closest integer $\leq x$. From this expression one can
derive important consequences for the 4d irreducible (i.e.~nonabelian)
anomaly. We have shown that, once the 6d irreducible bulk anomaly is
canceled, the first term does not contribute to the nonabelian 4d
anomaly.  On the other hand, the last term might be canceled by a 6d
Green-Schwarz four-form (or its dual, an axion).  Of course, it can
also be canceled amongst fermions, for instance if a vector-like 6d
fermion content is chosen.  The term is nonzero only for $\mathcal
G=SU(N)$ and vanishes even there if the fermion transforms in a real
representations.  The second term is a purely four-dimensional
irreducible anomaly that can only be removed with the introduction of
appropriate localized fermions at the singularity. Contrary to the
other two terms, its coefficient is an integer, which is a necessary
condition for local anomaly cancellation to occur. We have presented
numerous examples in the context of supersymmetric orbifold GUTs.

We have also calculated the localized anomaly of 6d Lorentz symmetry
and general covariance. To this end we have identified the remnants of
the 6d gravitational symmetries at the 4d fixed point located at $y_f$
besides the corresponding 4d ones. In terms of the 6d parameters
$\Lambda_{AB}(x,y)$ and $\xi^M(x,y)$ they are generated by
$\Lambda_{ab}(x,y_f)$, $\partial_m \xi^m(x,y_f)$ and
$\partial_{[m}\xi_{n]}(x,y_f))$. We find that only the first and
second symmetries have localized anomalies.  While the scaling
symmetry $\partial_m \xi^m$ can be rendered consistent by adding
suitable local counterterms to the action, quantum consistency of the
remnant Lorentz symmetry $U(1)_L$ requires that localized fields are
charged under it. This is rather surprising from a field theory
point of view, but occurs naturally in string theory orbifolds. The
localized fields of the low energy limit correspond to zero modes of
twisted strings. The latter do in fact transform non-trivially under
$U(1)_L$ (or, more general under $SO(6)$) and hence are expected to
contribute to the anomaly.

Let us conclude by pointing out some interesting directions of further
research. Clearly, most of our findings can more or less directly be
generalized to any codimension-two orbifold-singularity. Of course,
the structure of the $d-2$ dimensional anomalies is different, but our
general findings for the gauge anomalies will not change. Another
interesting direction concerns codimension-two conical singularities
of opening angles other than $2 \pi/n$ and gauge twists whose order is
unrelated to the geometric twist~\cite{Hebecker:2003jt}. The fact that
our results are independent of $n$ and only depend on the
$\delta$-function like background flux $\vev F$ points towards similar
results in these cases. On the technical side however, our calculation
crucially depended on the ability to define our theory on a simply
connected smooth covering space (with some constraints on the fields)
which is no longer possible, even for rational opening angles such as
$2\pi p/q$. Other, more refined tools are necessary, for instance the
form of the heat kernel on arbitrary conical singularities (see
Ref.~\cite{Dowker:1977zj}).  The third obvious generalization concerns
orbifold singularities of codimension other than two. Although higher
dimensional analogues of the finite sums in App.~\ref{evalsums} can be
performed and yield similar polynomials in the logarithm of the
orbifold twist, the immediate consequences for anomaly cancellation
tend to be more involved than in the codimension-two case.  The
remnant local Lorentz symmetry at the singularity becomes a nonabelian
one and one expects more stringent restrictions on the localized
matter content to achieve anomaly cancellation.

\section*{Acknowledgments}
I would like to thank A.~Hebecker, C. Scrucca, M. Serone and
R.~Sundrum for discussions. This work was supported by grants
NSF-PHY-0401513, DE-FG02-03ER41271 and the Leon Madansky Fellowship,
as well as the Johns Hopkins Theoretical Interdisciplinary Physics and
Astrophysics Center .

\appendix

\section{Notations and Conventions}
\label{conv}

We will adopt the follwoing index convention: lower case greek letters
take values $0\dots 3$, lower case latin letters $4\dots d-1$ and
capital latin letters take values $0..d-1$. Moreover, letters from the
beginning of the alphabet ($A,B,a,b,\alpha,\beta$) are flat space
(Lorentz) indices, while letters from the middle of the alphabet
($M,N,m,n,\mu,\nu$) denote curved (Einstein) indices.  Furthermore we
will denote the extra dimenisonal coordinates with $y$
(i.e. $y_1=x_4,\ y_2=x_5$) and often choose our coordinates such that
the fixed point under consideration lies at $y=0$.

We will work in the Euclidean. Our $d$ dimensional $\Gamma$ matrices
satisfy the Clifford algebra
\be
\{\Gamma^A,\Gamma^B\}=2\delta^{AB}\,.
\ee
In $d=6$ we write this algebra in terms of the four-dimensional
$\gamma$ matrices:
\be
\Gamma^a=\gamma^a\otimes\sigma^3,\qquad \Gamma^4=1\otimes \sigma^1,\qquad \Gamma^5=1\otimes \sigma^2\,,
\ee
and we apply this construction recursively for higher dimensions. 
The $SO(d)$ generators are given by 
\be
J^{AB}=\frac{i}{4}[\Gamma^A,\Gamma^B]\,.
\label{Lorentz}
\ee
The four-dimensional chirality matrix $\gamma^5$ is defined as
$\gamma^5=\gamma^0\cdots\gamma^3=\operatorname{diag}(+1,+1,-1,-1)$. The
$d$ dimensional chirality matrix (the analogue of $\gamma^5$) is
defined as
\be
\Gamma=\gamma^5\otimes\sigma^3\otimes\sigma^3\otimes\dots
\ee
The convention for the totally antisymmetric $SO(d)$ tensor is
$\epsilon^{0\cdots d-1}=+1$

The particular choice for the $\Gamma$-matrices leads to a trivial
embedding of $SO(4)$ spinor representations into $SO(d)$ ones:
\be
J^{\alpha\beta}_\Gamma=
\left( 
\begin{array}{ccc}
J_\gamma^{\alpha\beta}&&\\
&J_\gamma^{\alpha\beta}&\\
&&\ddots
\end{array} \right)\,,
\ee
where the subscript indicates the Clifford algebra matrices used to
compute Lorentz-generators according to Eq.~(\ref{Lorentz}).  
Finally, in $d=6$ the generator for rotation in the $45$-plane is given by
\be
\frac{1}{2}\Lambda^{ab}J_{ab}=\Lambda^{45}J_{45}=-\Lambda^{45}\frac{\sigma_3}{2}\,.
\ee
where $\Lambda_{45}$ is the angle of rotation.

\section{The Evaluation of the Heat Kernel}
\label{heatkernel}

The object formally defined as
\be
K(x,x',s)\equiv
\langle x |\exp(- s \Delta )| x'\rangle
\ee
is called the Heat Kernel of the Laplace type differential operator
$\Delta $. 
It satisfies the heat equation (hence the name)
\be
\frac{\partial}{\partial s} K (x,x',s)= -\Delta K(x,x',s)\,,
\label{heat}
\ee
obeying the boundary condition
\be
K(x'x',0)=\delta(x,x')\,,
\label{bc}
\ee
where $\delta(x,x')$ denotes the delta funcion (density) in curved
space. The heat kernel appears in the calculation of anomalies,
Eq.~(\ref{master}) with $\Delta$ the square of the Dirac operator,
\be
\Delta=-D^2\!\!\!\!\!\!/\ = -D^2 +E\,,\qquad
E=-\frac{1}{4} R +\frac{i}{2}\Gamma^{MN} F_{MN}\,.
\label{Dirac}
\ee
The covariant derivative on the right hand side of Eq.~(\ref{Dirac})
contains spin, Christoffel and gauge connections, $F_{MN}$ stands
short for $F_{MN}^a T_a$ with the hermitian generators $T_a$ and $R$
is the scalar curvature.
Follwing deWitt \cite{:1985bt} we write $K(x,x',s)$ as a power series
in the parameter $s$:
\be
K(x,x',s)\equiv(4 \pi s)^{-d/2} D^{1/2}(x,x') \exp(-\sigma(x,x')/2s 
)
\sum_{r\geq 0} a_r (x,x') s^r\,.
\label{ansatz}
\ee
The quantity $\sigma(x,x')$ is called the geodesic biscalar and
satisfies the differential equation
\be
\frac{1}{2}\sigma_;^{M}\sigma_{;M}=\sigma\,,
\label{sigma}
\ee
where the semicolon denotes covariant differentiation w.r.t.~$x$. The correspondig boundary condition reads
\be
[\sigma](x)\equiv\sigma(x,x)=0\,.
\ee
From here on the brackets denote the coincidence limit $x'=x$.
$\sigma$ is a biscalar function whose magnitude equals half the
geodesic distance squared between $x$ and $x'$. It is the
generalization of $(x-x')^2/2$ in flat space. The following
coincidence limits hold for $\sigma$
\be
[\sigma_{;M}]=0\,,\qquad [\sigma_{;MN}]=g_{MN}\,,\qquad[\sigma_{;MNR}]=0\,.
\ee
Further coincidence limits can be obtained by repeated covariant
differentiation of Eq.~(\ref{sigma}). The other quantity appearing in the ansatz Eq.~(\ref{ansatz}) is the van Vleck-Morette determinant defined 
%
%
by the differential equation
\be
D^{-1}(D{\sigma_;}^M)_{;M}= d\,,
\ee
together with the boundary condition
\be
[D]=\det g_{MN}\equiv g\,.
\ee
The normalization in Eq.~(\ref{ansatz}) as well as the limit $[a_0]=1$
are chosen such as to correctly reproduce the boundary condition,
Eq.~(\ref{bc}). In other words
\be
\lim_{s\to 0} 
\biggl\{
(4 \pi s)^{d/2} D(x,x')^{1/2} \exp(-\sigma(x,x')/2s )
\biggr\}
=\delta(x,x')\,.
\ee
Inserting the ansatz Eq.~(\ref{ansatz}) in the heat equation
(\ref{heat}), one obtains the recursion relations
\be
{\sigma_;}^Ma_{0;M}=0\,,
\label{rec1}
\ee
\be
{\sigma_;}^Ma_{r;M}+r\,a_r=
D^{-1/2}{(D^{1/2}a_{r-1})_{;M}}^M -E\,a_{r-1}\,.
\label{rec2}
\ee
From these relations and their covariant derivatives one can obtain
the coincidence limits $[a_r]$ in a completely straightforward (but
for higher $r$ increasingly tedious) manner.  The $[a_r]$ are local
covariant functionals of the curvature and field strength tensors and
of their covariant derivatives (for details on the calculation see
Ref.~\cite{:1985bt}).

To find the anomaly, Eq.~(\ref{master}), we need to calculate
\be
\lim_{s\to 0}\lim_{x'\to x} Q(x,x'') K(x'', x',s)\,.
\ee
As mentioned in Sec.~\ref{anomalies}, the term proportional to the
identity in $Q$ gives rise to the one $n$th the $d$ dimensional bulk
anomaly,\footnote{The factor of $1/n$ accounts for the fact that we
are essentially working on the covering space of the orbifold -
restricting to the fundamental domain, which equals again one $n$th
the volume of the covering space, compensates this factor and one
obtains precisely the $d$ dimensional result.} while the other terms
produce brane anomalies. They are proportional to
\be
K^{(k)}(x)\equiv\lim_{s\to0}\lim_{x'\to P^k x}K(x,x',s)\,.
\label{Korb}
\ee
Clearly, unless $x=Px$, that is at the fixed point $y=0$, this
quantity is exponentially suppressed in the limit $s\to 0$ while it
diverges for $y=0$. We thus expect $K^{(k)}(x)$ to be a distribution
centered at $y=0$.  Indeed, one can calculate
\be 
\lim_{s\to 0}\ \left(\frac{D (x,P^kx)}{(4\pi s)^{m} }\right)^{\frac{1}{2}}
\exp(-\sigma(x,P^k x)/2s) = \tilde g^{1/2}\,
\nu_k^{-1}\,
\delta(y)  \,,
\ee
where we have defined $\nu_k=\det(1-P^k_*)$, 
$P_*$ being the derivative of the orbifold symmetry
transformation $P$,\footnote{$P_*$ is here restricted to the $m$
dimensional subspace it acts upon non-trivially.} and $\tilde g$ denotes
the induced $d$ dimensional metric at the fixed point. 

For the evaluation of $K(x)$, note that all terms in the sum over $r$
with $r>r_0=(d-m)/2$ vanish in the limit $s\to 0$.  The remaining
terms greatly simplify when the trace over the gamma matrices is
taken in Eq.~(\ref{master}).  Notice that in our conventions
(see App.~\ref{conv}) the $d$-dimensional chirality matrix $\Gamma$ is
nothing but the tensor product of the $d-m$ dimensional one
$\tilde\Gamma$ and $m/2$ factors of $\sigma^3$. In fact, one can show
(in complete analogy to the usual argument for anomalies on
nonsingular spaces) that the coefficients $a_r$ with $r<r_0$ do not
contribute, as they do not provide sufficient number of $\Gamma^\mu$
matrices in order to ``saturate'' $\tilde \Gamma$. We are left with
precisely the $s$-independent term
\be
K^{(k)}(x)=\frac{1}{(4\pi)^{r_0}\,
\nu_k}\,
\tilde g^{1/2}
\delta(y)\,
[a_{r_0}]\,.
\label{sindependent}
\ee
Let us now specialize to the case $d=6$, $m=2$.  The heat kernel
coefficient $[a_2]$ can be evaluated by use of the recursion relations
Eq.~(\ref{rec1}) and (\ref{rec2}).\footnote{For details on this see
e.g.~Refs.~\cite{:1985bt}. Again, great simplifications occur as terms
with less than four $\Gamma^\alpha$ can be discarded.} One finds
\be
[a_2]=\frac{1}{8}\Gamma^{\alpha\beta}\Gamma^{\gamma\delta}\left(\frac{1}{24}R_{\alpha\beta MN}{R_{\gamma\delta}}^{MN}-F_{\alpha\beta} F_ {\gamma\delta}
\right)
+\dots
\ee
where the dots denote terms with a lower number of
$\Gamma^\alpha$'s. Finally we insert the explicit expressions for
$\Gamma$, $P_L$ and $P_*$
\be
\Gamma=\gamma^5\otimes\sigma^3\,,\qquad P_L=\exp( -i \pi/ n \sigma^3)\,,
\qquad \nu_k=4\sin^2(\pi k/n)  \,,
\ee
to obtain the final expression for the localized anomaly
\be
\mathcal A(x)=-\frac{i}{8 \pi^2 n}\sum_k \frac{1}{\sin(\pi k/n)} 
\Tr (\Lambda P^k_{\mathcal G} A_2)
\delta(y)\,,
\ee
where the trace is now over the gauge degrees of freedom only and the
quantity $A_2$ is given by
\be
A_2=\frac{1}{4}\Tr \gamma^5 [a_2]
=\frac{1}{8}\epsilon^{\alpha\beta\gamma\delta}\left(\frac{1}{24}R_{\alpha\beta MN}{R_{\gamma\delta}}^{MN}-F_{\alpha\beta} F_ {\gamma\delta}
\right)\,.
\label{A2}
\ee
Finally, for the gravitational anomaly we also need to evaluate
\be
K_m^{(k)}(x)\equiv
\lim_{x'\to P^k x}D_m K(x,x',s)\,.
\ee 
The term with $r=2$ in the expansion Eq.~(\ref{ansatz}) contributes
\be
\frac{1}{(4\pi)^2}\nu_k^{-1}
\biggl\{
\delta(y)_{;m}[a_2]+
\delta(y)[a_{2;m}]
\biggr\}\,.
\ee
However, there is a subtlety involved in this case: The term with
$r=1$ also contributes, as can be seen from performing a covariant
Taylor expansion\footnote{See for instance
Ref.~\cite{Avramidi:2001ns}.} around the point $x'$:
\be
a_1(x,x')=[a_1](x')+\sigma^N [a_{1;N}](x')+\frac{1}{2}
\sigma^N\sigma^M[a_{1;(NM)}](x')+
\dots\,,
\ee
where $\sigma^M={\sigma_;}^M(x,x')$ and the dots stand for terms of
higher order in $\sigma^M$. The second order term in this expansion,
combined with the exponential in Eq.~(\ref{ansatz}), gives indeed an $s$
independent contribution
\be
\frac{1}{(4\pi)^2}\nu_k^{-1}
{\delta(y)_;}^{n}[{a_{1;(mn)}}]\,.
\ee
From the recursion relations Eq.~(\ref{rec1}) and (\ref{rec2}) one
easily obtains
\be
[{a_{1;(mn)}}]=\frac{1}{96}
\Gamma^{\alpha\beta}\Gamma^{\gamma\delta} {R_{\alpha\beta sm}}{{R_{\gamma\delta }}^s}_n+\dots
\ee
However, this tensor is not only symmetric w.r.t.~$m$ and $n$, but it is even proportional to the identity,
\be
[a_{1;(mn)}]=
\frac{1}{2} [{a_{1;\,s}}^s ]g_{mn}\,.
\ee
The result for $K^{(k)}_m(x)$ can thus be written as
\be
K^{(k)}_m(x)=
\frac{1}{(4\pi)^2\nu_k}\tilde g^{1/2}
\biggl(
\delta(y)_{;m}
\bigr\{[{a_{1;s}}^s]+
[a_2]\bigl\}
\  +\ \delta(y)[a_{2;m}]
\biggr)\,.
\ee
The last term will not contribute to the localized anomaly, as
$\xi^m$ vanishes at the fixed point. 

\section{Evaluation of the Sums}
\label{evalsums}

We would like to compute
\be
X_n(q)=\sum_{k=1}^{n-1} \frac{q^k}{\sin(\pi k/n)}\,,\qquad q^n=-1\,.
\ee
Using the identity
\be
\frac{1}{\sin(\pi x)}=
\frac{1}{\pi}\sum_{\ell=-\infty}^{+\infty}\frac{(-)^\ell}{x-\ell}
\ee
we rewrite this as
\be
X_n(q)=\frac{n}{\pi}\sum_{\ell,k}\frac{q^{k-n\ell}}{k-n\ell}\,,
\ee
where we have made use of the fact that $q^n=-1$. By inspection of the
possible values for ${\ell,k}$ we observe that the double sum can be
written as a single one
\be
X_n(q)=\frac{n}{\pi}\sum_{\ell'\neq 0}\frac{q^{\ell'}}{\ell'}\,.
\ee
Observing that $\sum(q^\ell/\ell)=-\log(1-q)$ with the branch cut
going along the negative real axis we obtain
\be
X_n(q) =\frac{n}{\pi}\left(\log(1-\bar q)-\log(1- q)\right)
=-\frac{n}{\pi} \log (-q)
=\frac{n}{\pi} (i\pi-\log' q) \,,
\label{sum1}
\ee
where by definition the branch cut of $\log'$ is going from 0 to $(+\infty-i\epsilon)$.

Next consider the sum
\be
X^{(2)}_n(q)=\sum_{k=1}^{n-1} \frac{q^k}{\sin(\pi k/n)^2},\qquad q^n=+1\,.
\ee
This time the identity
\be
\frac{1}{\sin(\pi x)^2}=\frac{1}{\pi^2}\sum_{\ell=-\infty}^{+\infty} \frac{1}{(\ell-x)^2}
\ee
is used to rewrite the sum as
\be
X^{(2)}_n(q)=\frac{n^2}{\pi^2}\sum_{k,\ell} \frac{q^{k-n\ell}}{(k-n\ell )^2}\,,
\ee
where again $q^n=1$ has been used. As before this can be combined into
a single sum:
\be
X^{(2)}_n(q)=\frac{n^2}{\pi^2}\sum_{\ell'\neq0}\left( \frac{q^{\ell'}}{(\ell' )^2} 
-\frac{1}{n^2\ell'^2}\right)\,.
\ee
This is in fact nothing but the definition of the poly logarithm functions
\be
X_n^{(2)}(q)=\frac{n^2}{\pi^2}\left(\Li_2(q)+\Li_2(\bar q)-
\frac{2}{n^2}\zeta(2)\right)\,.
\ee
Using the relation of $\Li_2$ to the Bernoulli polynomials as well as
$\zeta(2)=\pi^2/6$ we find
\be
X_n^{(2)}(q)=- 2n^2 \left(\frac{\log^2(-q)}{4\pi^2}+
\frac{1}{12}+\frac{1}{6 n^2}\right)\,.
\ee
Note that this result implies
\be
\frac{1}{2n} \sum_{k=1}^{n-1}\frac{\cos(\pi k/n)}{\sin^2(\pi k/n)} q^k
=
n \left(\frac{\log (-q)}{2\pi i}\right)^2
+\frac{1}{12}\left(\frac{1}{n}-n\right)\,,
\label{sum3}
\ee 
for $q^n=-1$.  This result is needed for the evaluation of the anomaly
of the remnant Lorentz symmetry in Sec.~\ref{grav}.

\end{document}